\documentclass[12pt]{elsart}

\usepackage{graphicx}
\usepackage{subfigure}
\usepackage{amsmath}
\usepackage{amssymb}
\usepackage{cite}

\journal{Physics Letters A}

\begin{document}

\begin{frontmatter}

\title{Magnetocaloric properties of frustrated tetrahedra-based spin nanoclusters}
\author{M. Mohylna},
\author{M. \v{Z}ukovi\v{c}\corauthref{cor}}
\ead{milan.zukovic@upjs.sk}
\address{Institute of Physics, Faculty of Science, P.J. \v{S}af\'arik University\\ Park Angelinum 9, 041 54 Ko\v{s}ice, Slovakia}
\corauth[cor]{Corresponding author.}

\begin{abstract}
Magnetization, entropy and magnetocaloric properties of various geometrically frustrated tetrahedra-based Ising antiferromagnetic nanoclusters with corner-, edge-, and face-sharing topologies are studied by exact enumeration. It is found that the studied properties strongly depend on the nanocluster topology and can be very different from those of a single tetrahedron as well as the pyrochlore lattice formed by an infinite number of corner-sharing tetrahedra. From the magnetocaloric point of view an important difference from the latter two systems is the absence of the ground-state zero-magnetization plateau in most of the studied structures, which at low temperatures facilitates emergence of a giant magnetocaloric effect in a vanishing magnetic field in the adiabatic demagnetization process. Magnetic systems with such properties might be suitable candidates for technological application as efficient refrigerators to ultra-low temperatures. 
\end{abstract}


\end{frontmatter}

\section{Introduction} 

Small spin clusters, magnetically isolated from the environment, can be viewed as building blocks of molecular nanomagnets, which have attracted considerable attention~\cite{kahn93,atta06,furr13} due to their natural occurrence in many real materials as well as the possibility of their artificial design in a highly controlled manner~\cite{shen02,loun07,pede14}. Their potential applications include data storage, quantum computing and molecule-based spintronics devices but they have also proved to be good candidates for a magnetic cooling technology as they can display a very large
magnetocaloric effect (MCE) at low temperatures~\cite{spic01,evan06,mano07,mano08}. MCE in high-spin molecular nanomagnets typically results from excesses of magnetic entropy at low temperatures due to macroscopic degeneracies of the molecular spin states: the higher the spin value the larger the magnetic entropy~\cite{evan05,shaw07}. Spin degeneracy can be further increased by designing very weak magnetic links between the single-ion spin centers~\cite{evan09}.\\
\hspace*{5mm} Another class of magnetic systems showing a high degeneracy at low temperatures are geometrically frustrated magnets. They can remain disordered and possess finite entropy even in ground state and thus the lowest achievable temperatures are not limited by the transition temperature as for example in paramagnetic salts. The latter are the standard refrigerant materials for magnetic cooling, however, their massive drawback is the limitation of the lowest achievable temperature by the temperature of a phase transition to diamagnetic or ferromagnetic state and the linear dependency between temperature and applied magnetic field. On the other hand, in geometrically frustrated antiferromagnetic systems the rate of the temperature decrease due to the varying of the external magnetic field is considerably bigger then the one characteristic for the paramagnetic salts and its variation is more complex. Indeed, it has been shown that in terms of MCE the field-dependent efficiency of a geometrically frustrated magnet can exceed that of an ideal paramagnet with equivalent spin by more than an order of magnitude~\cite{zhit03,schn07}. This enhancement is related to the presence of a macroscopic number of soft modes associated with geometrical frustration below the saturation field.\\
\hspace*{5mm} Frustration-enhanced MCE is more likely to be observed in low-dimensional systems, such as the 2D kagome or triangular antiferromagnetic lattices, however, it can also be observed in 0D systems, i.e., geometrically frustrated molecular nanomagnets. In particular, recent magnetocaloric measurements in ${\rm Gd}_7$ - a geometrically frustrated molecular nanomagnet - brought the first direct demonstrations of sub-Kelvin cooling ($\sim$ 200 mK) achieved with a molecular nanomagnet~\cite{shar14,pine16}. They also revealed isentropes with a rich structure, as a direct manifestation of the trigonal antiferromagnetic net structure. These results indicated the possibility to design and control the cooling power of molecular materials by selecting an appropriate topology of magnetic couplings between the interacting spins. Observations of the frustration-enhanced magnetocaloric effect in molecular nanomagnets triggered theoretical investigations of 0D systems of various shapes and topologies, such as geometrically frustrated Ising spin clusters with the shape of regular polyhedra (Platonic solids)~\cite{stre15,karl16} or composed of triangular units~\cite{milla1,zuko14,zuko15,zuko18}. The reported results pointed to great sensitivity of magnetocaloric properties to the spin cluster geometry. A giant magnetocaloric effect during the adiabatic demagnetization has been observed at the critical fields at which the magnetization jumps between different plateaux. Consequently, if efficient cooling to the lowest possible temperatures is targeted then the systems lacking zero-magnetization ground-state plateaux in an applied magnetic field, such as the Ising octahedron and dodecahedron from the Platonic solids~\cite{stre15}, the Star-of-David nanocluster~\cite{zuko18}, as well as most of the studied triangle-based clusters~\cite{zuko15}, appear to be promising candidates.\\
\hspace*{5mm} In the present study we focus on geometrically frustrated tetrahedra-based Ising nanoclusters. Corner-sharing tetrahedra are well known as building blocks of the pyrochlore structure. The latter has been intensively studied due to strong geometrical frustration resulting in the absence of long-range ordering and large residual entropy~\cite{hari97,rami99}, which gives such systems a great potential for showing an enhanced MCE and thus for being used for magnetic refrigeration at low temperatures through the adiabatic (de)magnetization processes. In particular, materials like $\rm{Gd}_2\rm{Ti}_2\rm{O}_7$~\cite{sosi05} or $\rm{Er}_2\rm{Ti}_2\rm{O}_7$~\cite{wolf16} appear promising for technological applications as efficient low-temperature magnetic refrigerators. In Ising-like pyrochlore structures the best cooling performance is achieved close to the critical fields at $h_{c1}=h_{sat}/3$ and $h_{c2}=h_{sat}$, where $h_{sat}$ is the saturation field~\cite{jurc17}, which are respectively associated with magnetization jumps between the zero to one-half and from one-half to the fully saturated plateaux~\cite{jurc14}. In the present approach we show that by a proper arrangement of tetrahedra in small clusters, formed by corner- , edge-, and face-sharing elementary units, one can design various highly degenerate structures mostly with no zero-magnetization plateaux, i.e., the first critical field $h_{c1}=0$, and thus achieve a giant MCE even close to zero magnetic field.

\section{Model}

\begin{figure}[t!]   
\centering 
\subfigure[2CS]{\includegraphics[scale=0.17,clip]{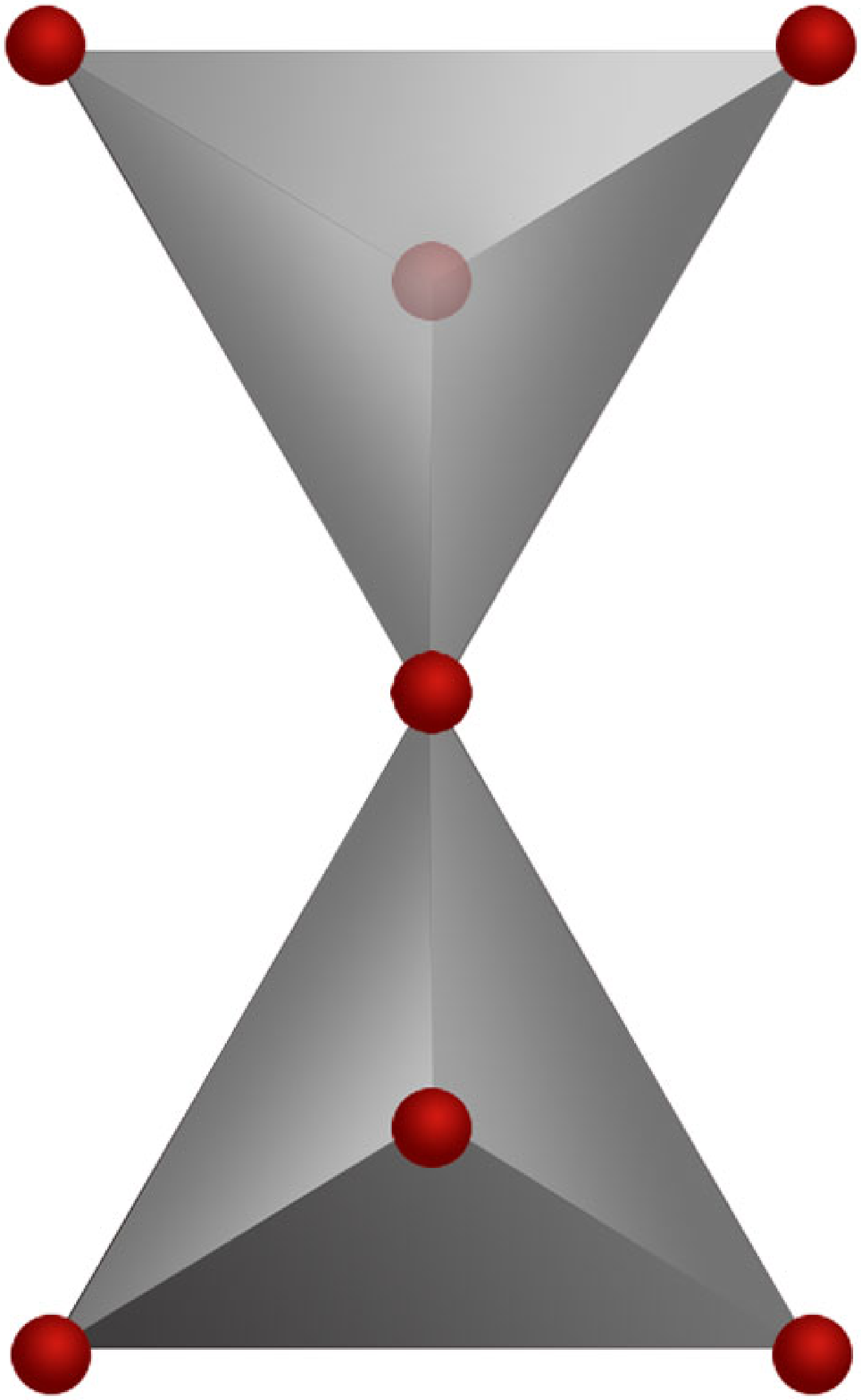}\label{fig:2cs}}  
\subfigure[2FS]{\includegraphics[scale=0.17,clip]{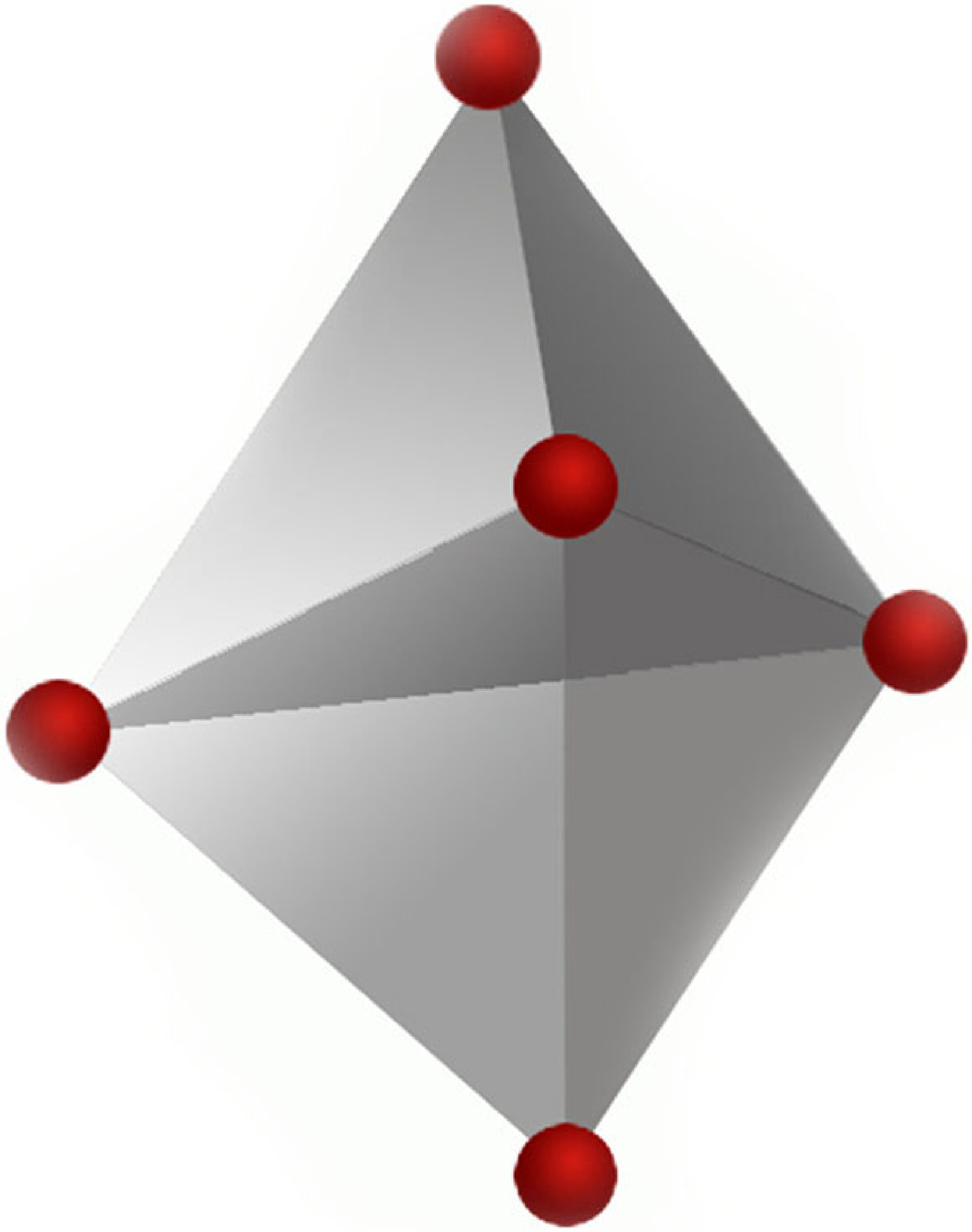}\label{fig:2fs}}
\subfigure[4CS]{\includegraphics[scale=0.17,clip]{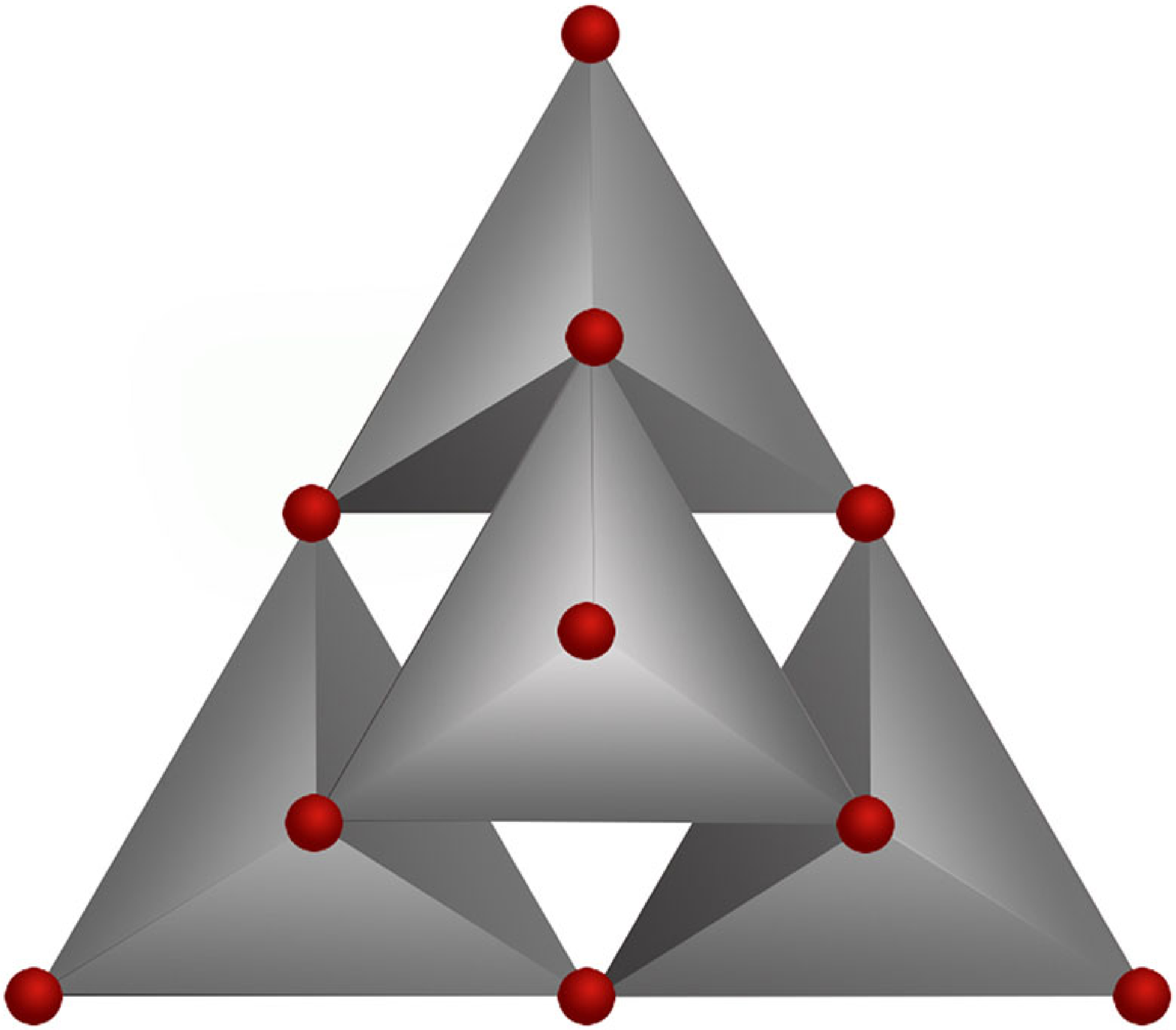}\label{fig:4cs}}\\
\subfigure[4ES]{\includegraphics[scale=0.17,clip]{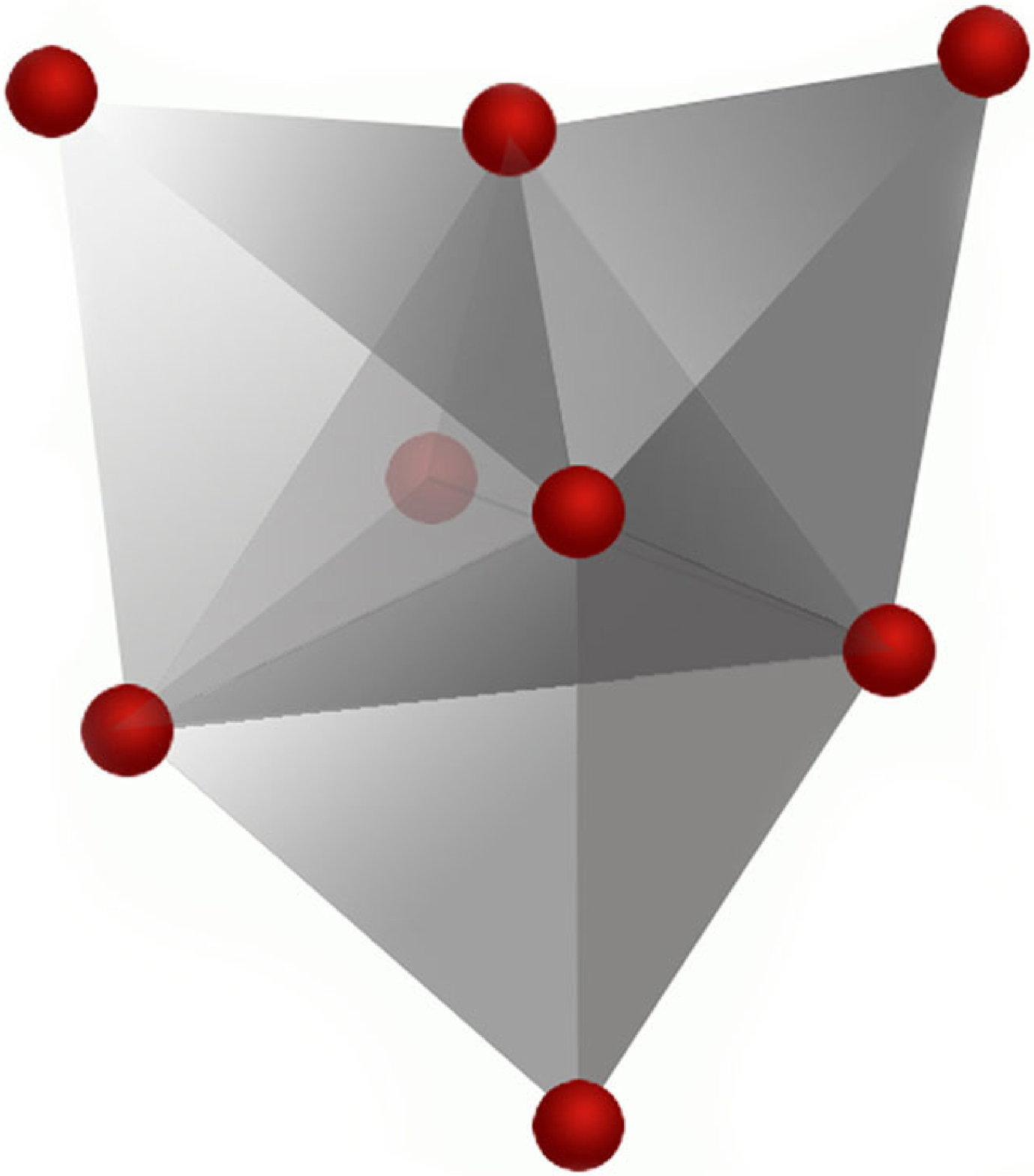}\label{fig:4es}}
\subfigure[5CS]{\includegraphics[scale=0.17,clip]{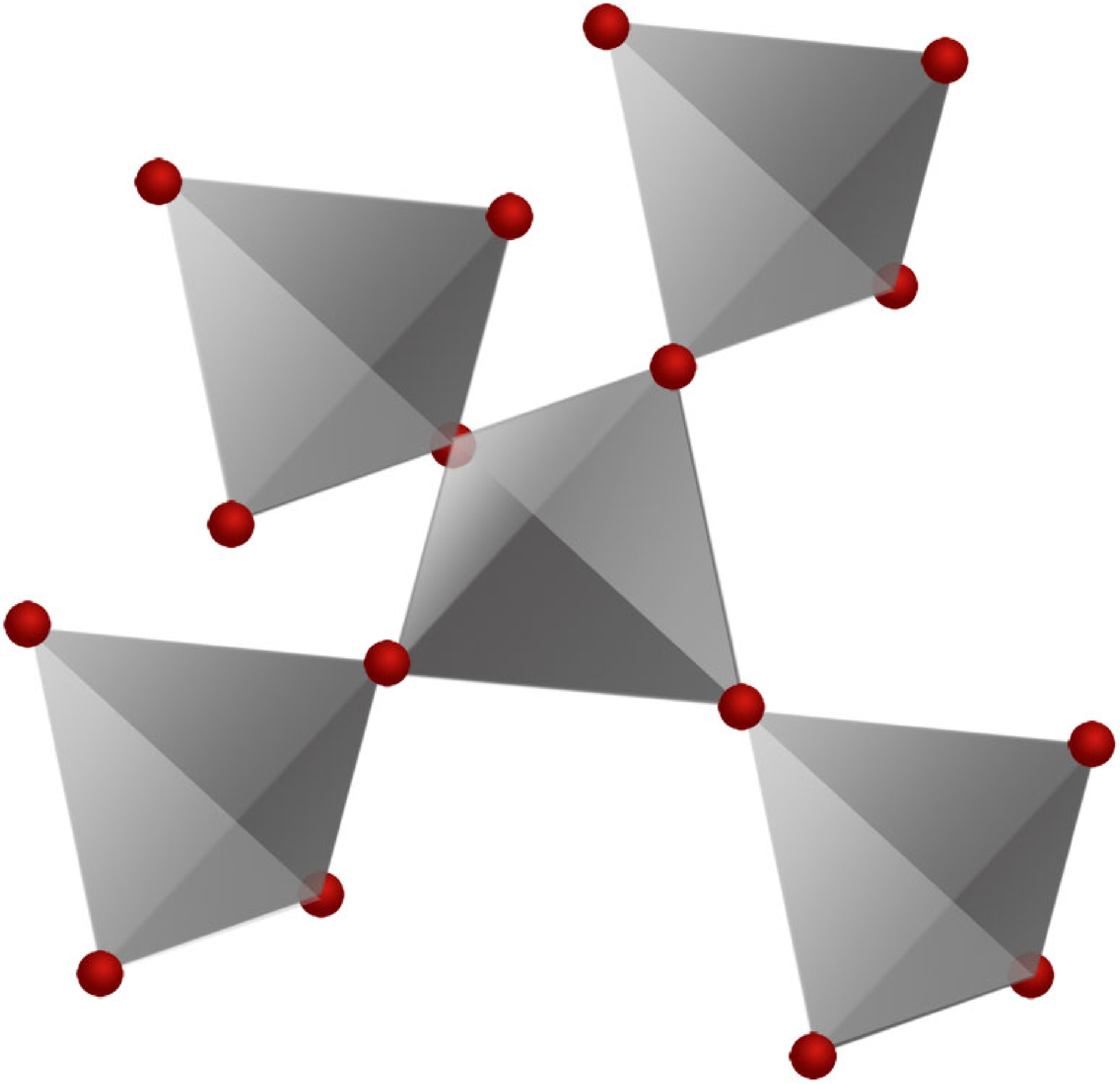}\label{fig:5cs}}
\subfigure[8ES]{\includegraphics[scale=0.17,clip]{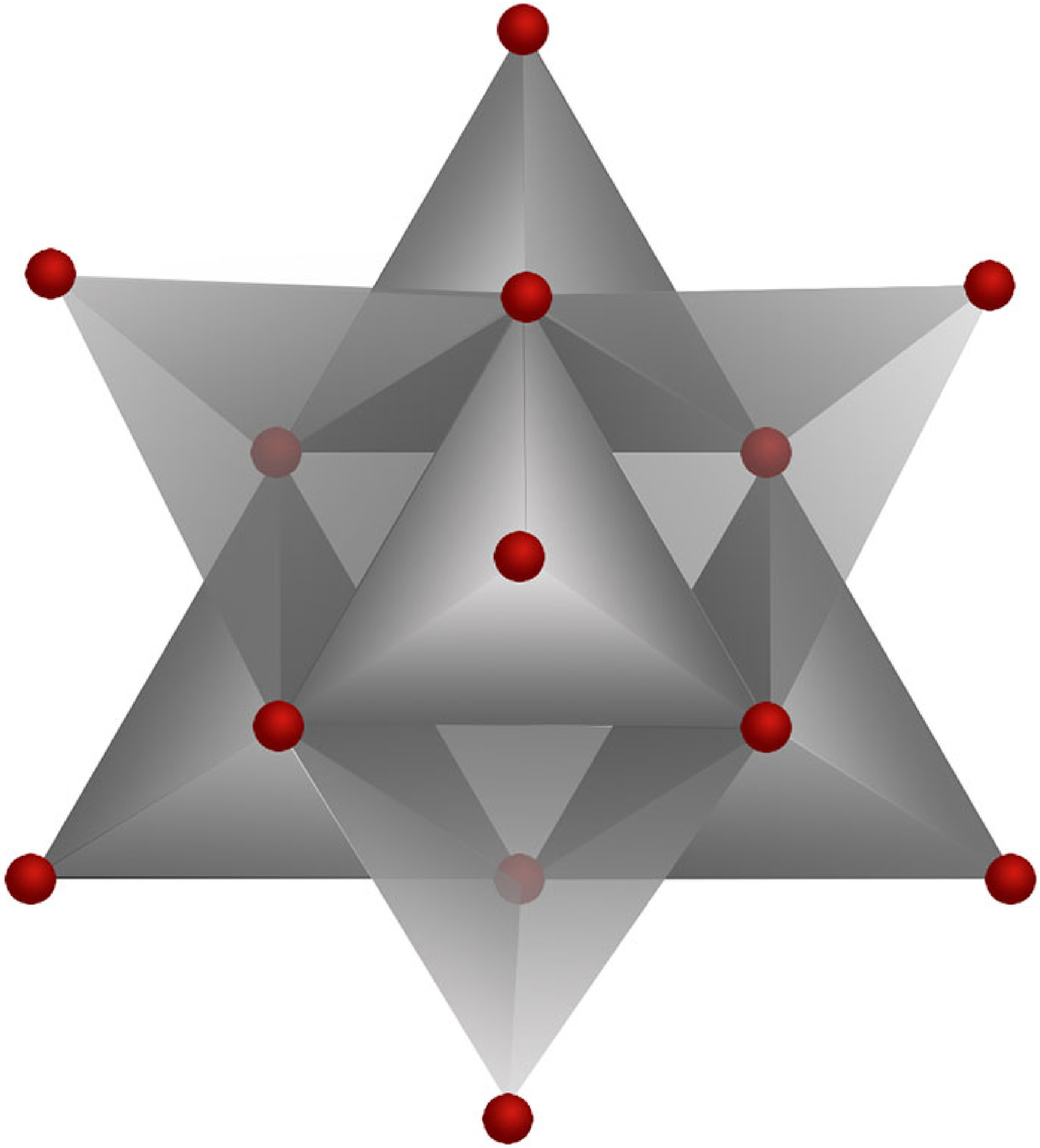}\label{fig:8es}} 
\caption{Considered spin clusters obtained by linking of $n=2,4,5$ and $8$ regular tetrahedra in (a,c,e) corner-sharing CS, (d,f) edge-sharing ES, and (b) face-sharing FS manners.} 
\label{fig:clusters}
\end{figure}

We consider the antiferromagnetic Ising spin system on small tetrahedra-based clusters of different topologies in an applied external magnetic field, described by the Hamiltonian 
\begin{equation} \label{Hamilt}
H=-J\sum_{\langle i,j\rangle}s_i s_j - h\sum_i s_i,
\end{equation}
where $s_i=\pm 1/2$ is the spin variable on the $i$-th site of a cluster, the summation $\langle i,j\rangle$ runs over all nearest neighbor pairs, $J<0$ is an antiferromagnetic exchange interaction (for simplicity hereafter we put $J=-1$), and $h$ is the external magnetic field. The considered spin clusters are presented in Fig.~\ref{fig:clusters}. They consist of a finite number of spins on frustrated structures designed by connecting elementary regular tetrahedra by vertices, edges or faces in the following ways: (a) 2CS: two corner-sharing, (b) 2FS: two face-sharing, (c) 4CS: four corner-sharing, (d) 4ES: four edge-sharing, (e) 5CS: five corner-sharing, and (f) 8ES: eight edge-sharing (stellated octahedron). The 8ES cluster can be formed from the 4CS cluster by attaching four tetrahedra on the triangular gaps on each face of the latter and thus in (f) one tetrahedron attached to the rear side is completely hidden. We note that exact results for a single tetrahedron were already reported in Ref.~\cite{stre15} and therefore we will not show them again here.
  
\section{Method}
\subsection{Ground state}
The relatively small system sizes consisting of maximum of $N=16$ spins allow direct enumeration of all the quantities of interest by scanning the configurational space and identifying those spin states $s_{\text{gs}}=\{s_1,s_2,\ldots,s_N\}$ which minimize the energy functional (\ref{Hamilt}) and thus correspond to the ground state (GS).\\
\hspace*{5mm} By counting the number of spin configurations corresponding to the minimal energy we calculate the GS degeneracy $W$. Magnetization of each identified GS configuration is then obtained as $M_{k} = \sum_{i=1}^N s_{\text{gs}}, k=1,2, \ldots W$. Putting the Boltzmann constant equal to one, we obtain the total magnetization per spin:
\begin{equation} \label{namtot}
m=\frac{1}{WN}\sum_{k=1}^W M_k
\end{equation}
and the entropy per spin
\begin{equation} \label{entr}
\frac{S}{N}=\frac{1}{N}\ln W.
\end{equation}

\subsection{Finite temperatures}
In order to extend our calculations to finite temperatures we calculate the density of states $g(M,E)$ as a function of the magnetization $M=\sum_{i=1}^N s_i$ and the exchange energy $E=\sum_{\langle i,j\rangle}s_i s_j$. Then the partition function can be obtained in the following form
\begin{equation} \label{stsum}
Z(T,h)=\sum_{M,E}g(M,E)e^{-\frac{E-hM}{T}}.
\end{equation}
Having obtained the partition function, a mean value of any thermodynamic function $A$ can be calculated as a function of the temperature and the field as
\begin{equation} \label{stred}
\langle A(T,h)\rangle =\frac{\sum_{M,E} A g(M,E)e^{-\frac{E-hM}{T}}}{Z(T,h)}.
\end{equation}
Using Eq.~(\ref{stred}) we obtain the magnetization per spin as 
\begin{equation} \label{magnsr}
m(T,h)=\frac{\langle M(T,h)\rangle}{N}
\end{equation} 
and the magnetic entropy density (entropy per spin) in the form
\begin{equation} \label{entrmain}
\frac{S(T,h)}{N}=\frac{U(T,h)-F(T,h)}{NT},
\end{equation}
where $U(T,h)=\langle E(T,h)-hM(T,h)\rangle$ is the enthalpy and $F(T,h)=-T \ln Z(T,h)$ is the free energy.\\
\hspace*{5mm} From the entropy as a function the temperature and the field, one can calculate quantities relevant for the study of magnetocaloric properties. Namely, the isothermal entropy density change is obtained as
\begin{equation} \label{entr_change}
\Delta S(T,\Delta h)/N =(S_f-S_i)/N, 
\end{equation}
where $S_i$ and $S_f$ are the entropies at the initial $h_i$ and final $h_f$ fields, respectively, and $\Delta h = h_f-h_i$. Another quantity of interest is the adiabatic magnetic cooling rate $C_r$ defined as
\begin{equation} \label{cool_rate}
C_r =-\frac{(\partial S/\partial h)_T}{(\partial S/\partial T)_h}, 
\end{equation}
where the lower indices $T$ and $h$ mean the constant temperature and field, respectively.

\section{Results} 
\subsection{Ground state}

\begin{figure}[t!]   
\centering 
\subfigure{\includegraphics[scale=0.32,clip]{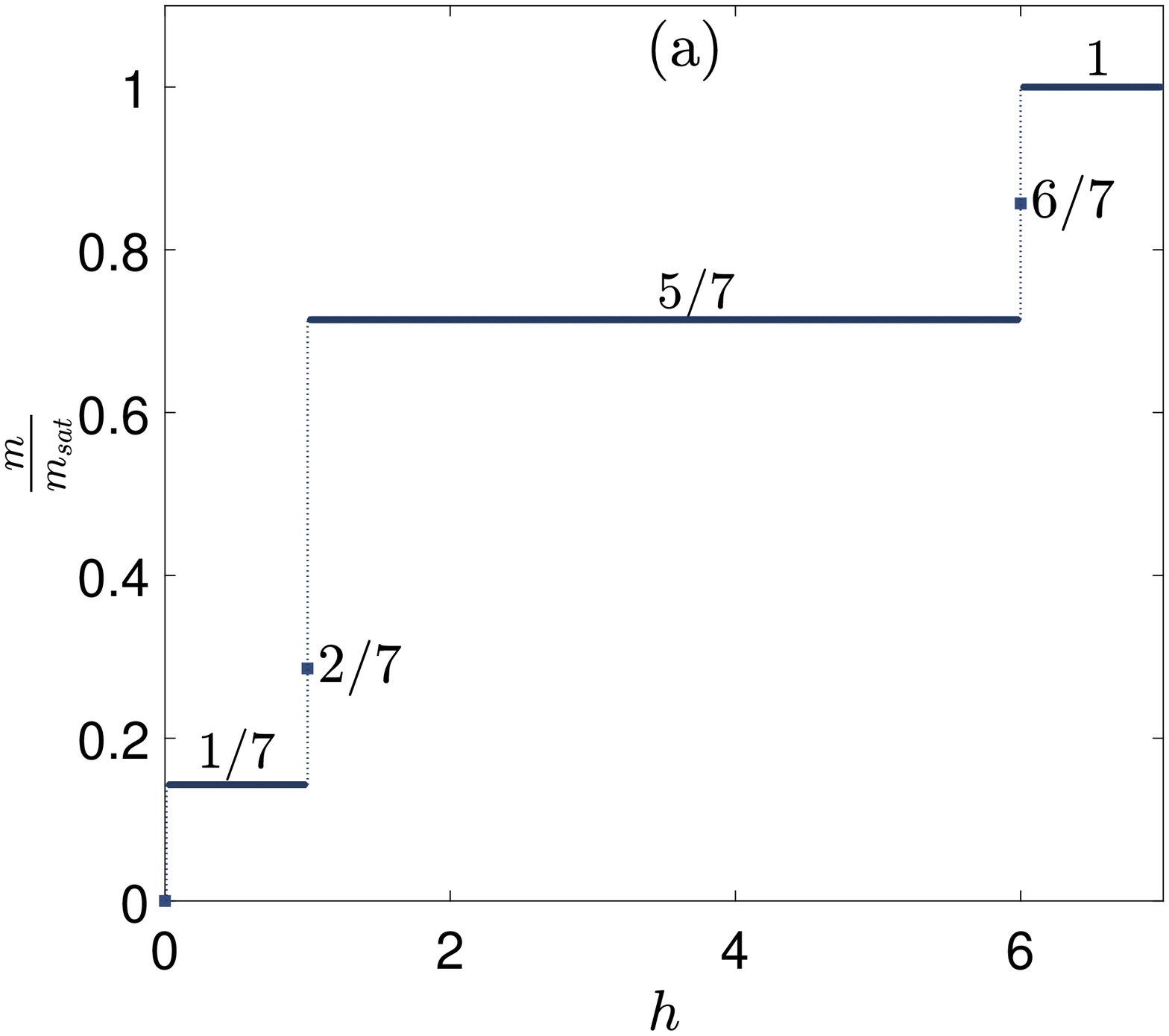}\label{fig:mag_gs_2cs}}  
\subfigure{\includegraphics[scale=0.32,clip]{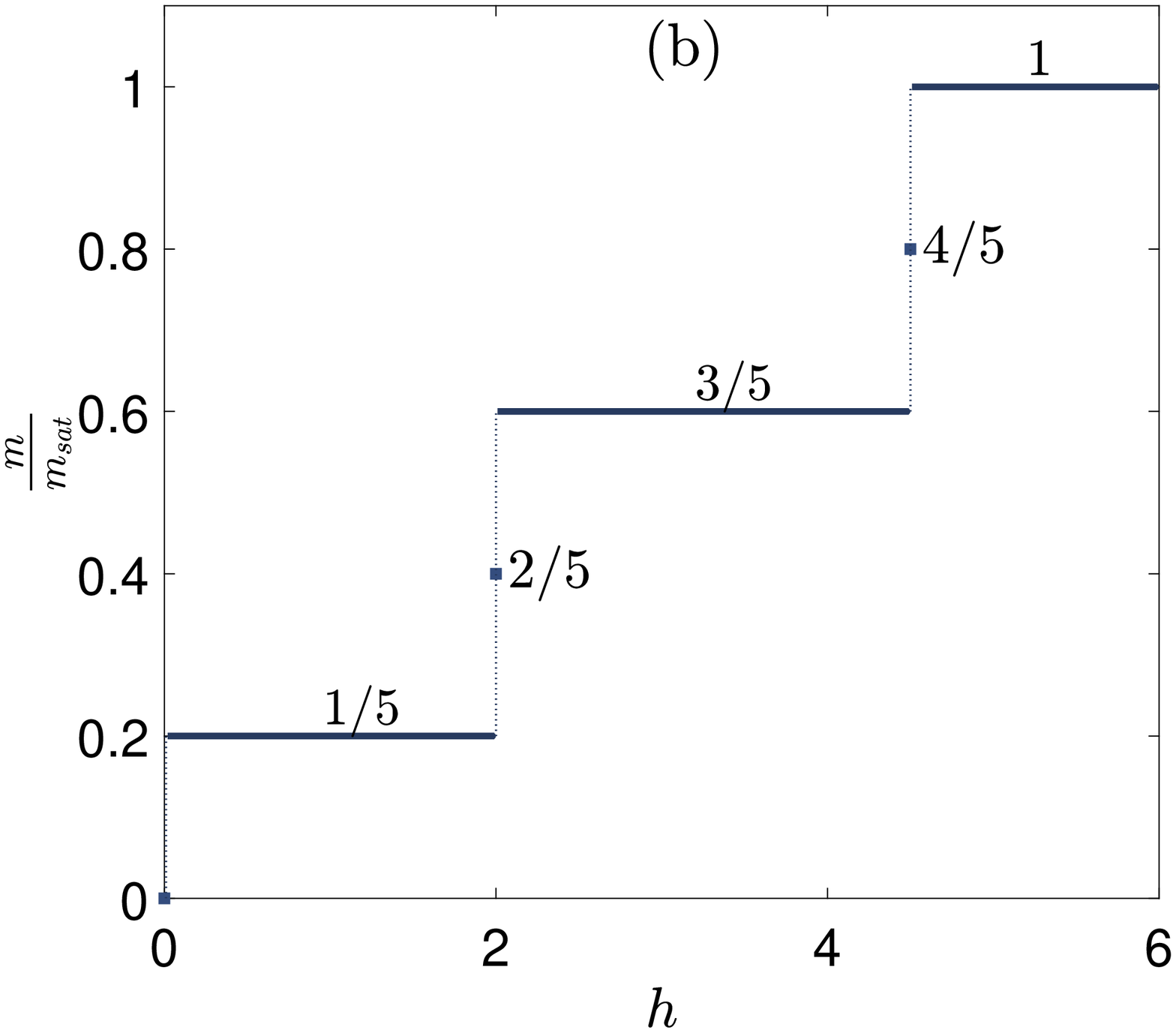}\label{fig:mag_gs_2fs}}\\
\subfigure{\includegraphics[scale=0.32,clip]{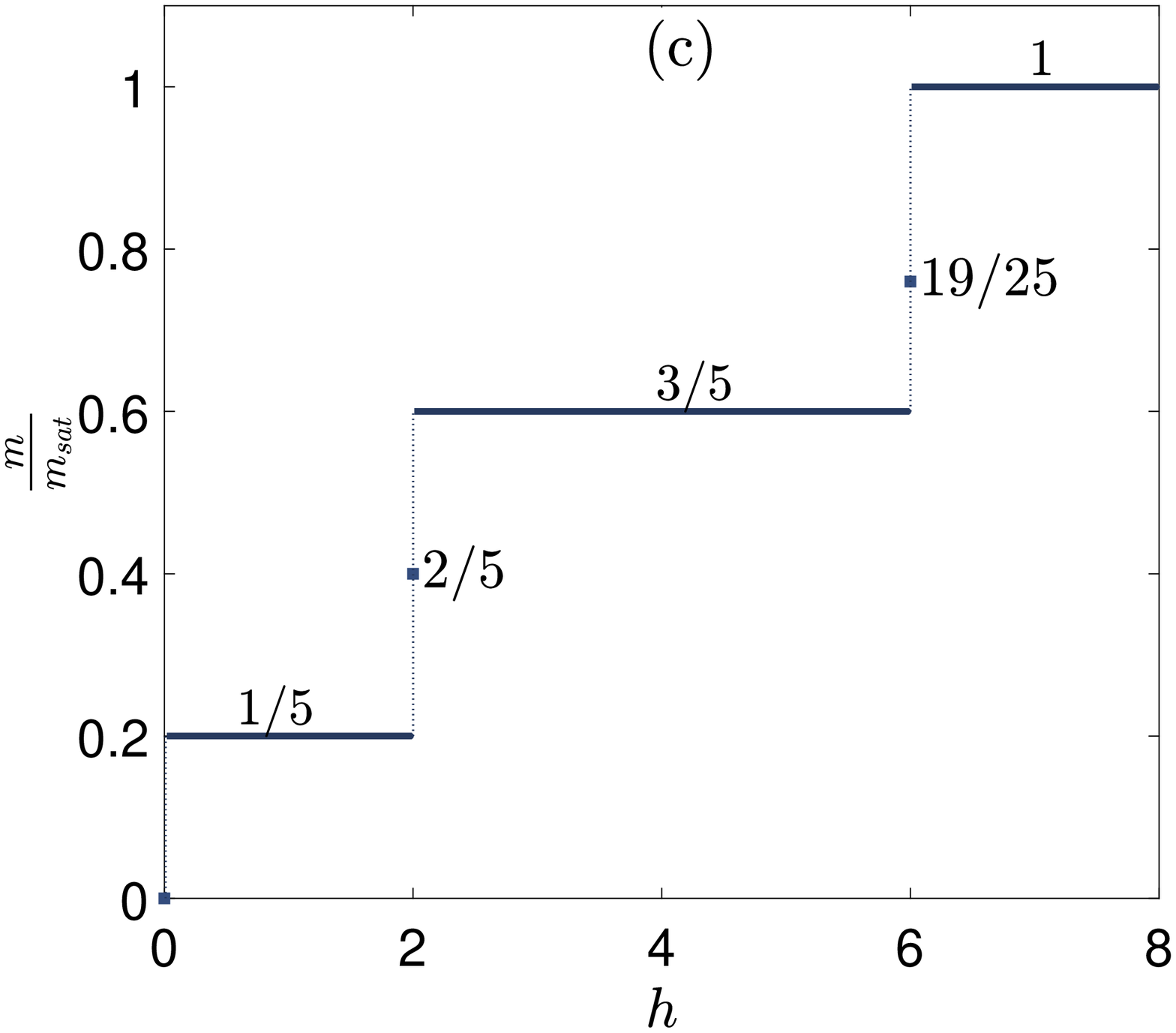}\label{fig:mag_gs_4cs}}
\subfigure{\includegraphics[scale=0.32,clip]{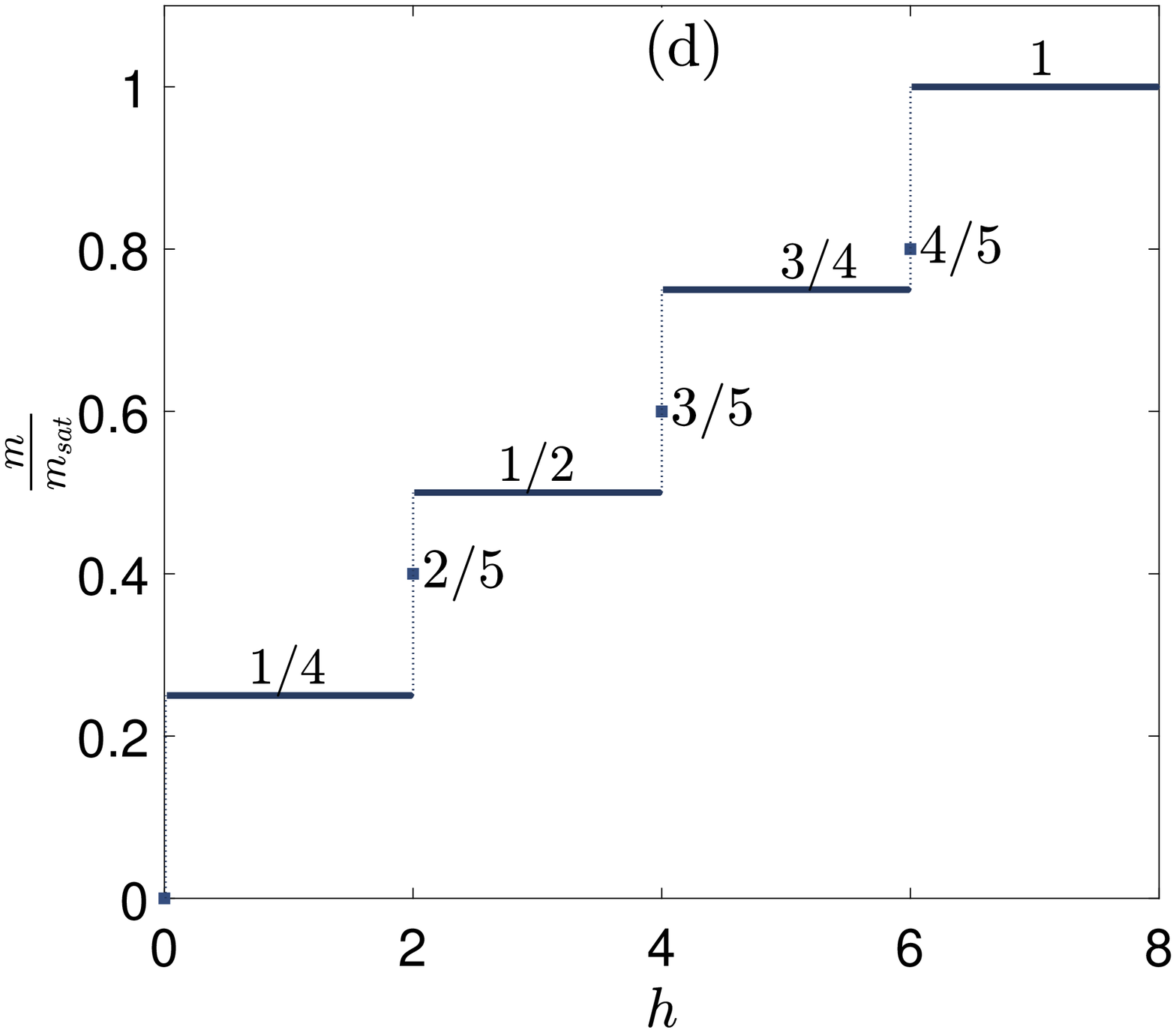}\label{fig:mag_gs_4es}}\\
\subfigure{\includegraphics[scale=0.32,clip]{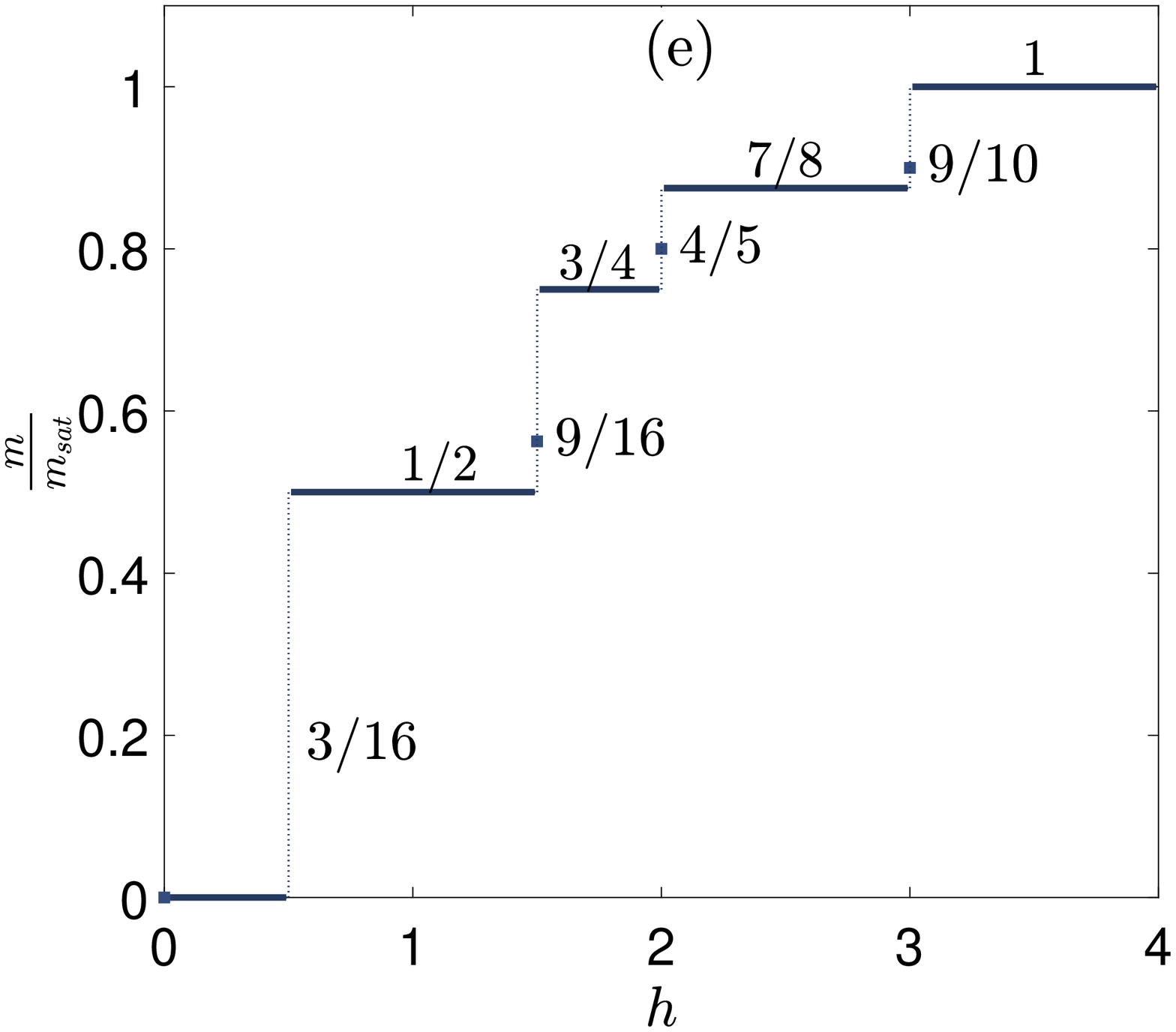}\label{fig:mag_gs_5cs}}
\subfigure{\includegraphics[scale=0.32,clip]{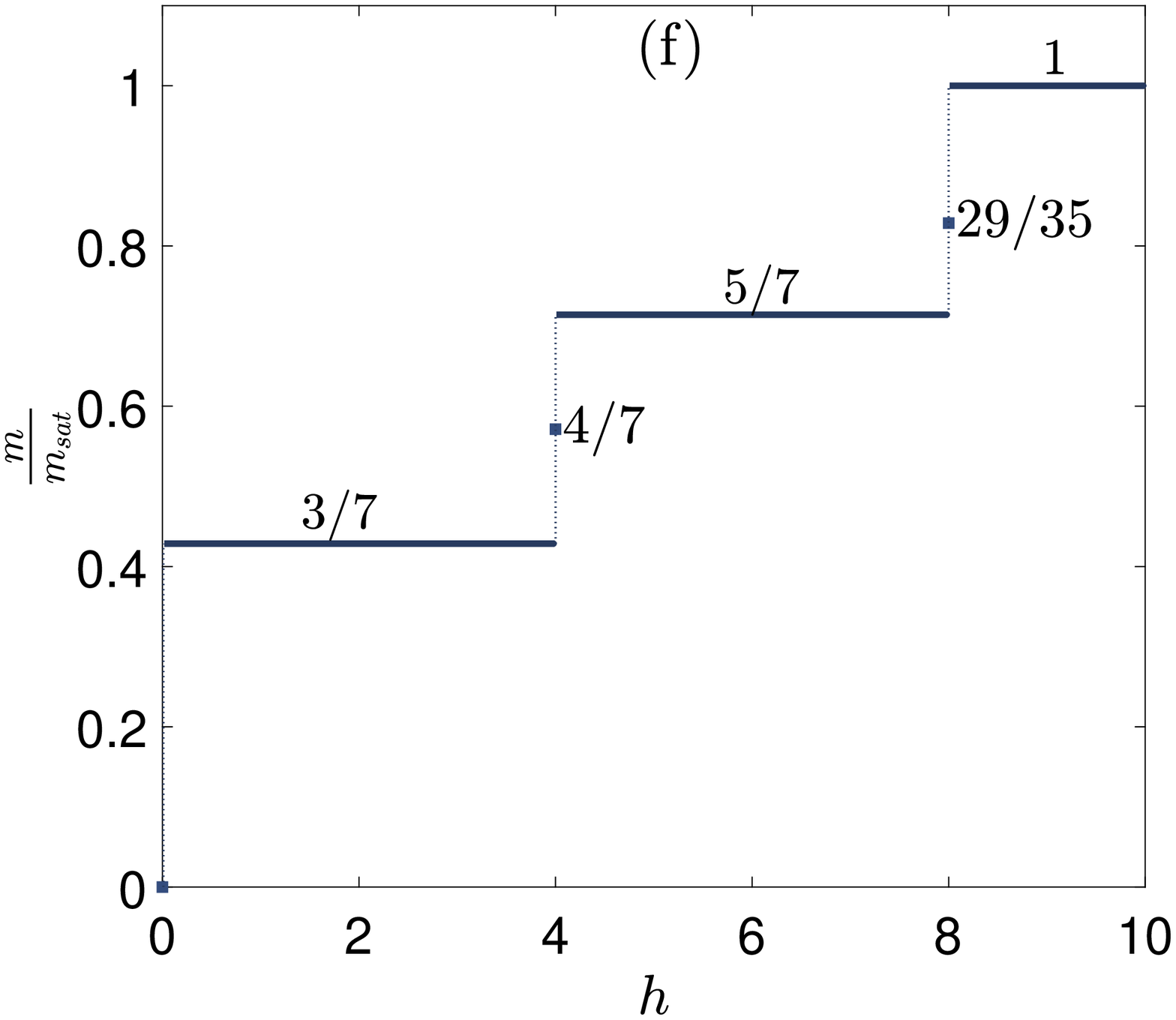}\label{fig:mag_gs_8es}} 
\caption{Field-dependence of the GS magnetization in (a) 2CS, (b) 2FS, (c) 4CS, (d) 4ES, (e) 5CS, and (f) 8ES clusters.} 
\label{fig:mag_gs}
\end{figure}

\begin{figure}[t!]   
\centering 
\subfigure{\includegraphics[scale=0.32,clip]{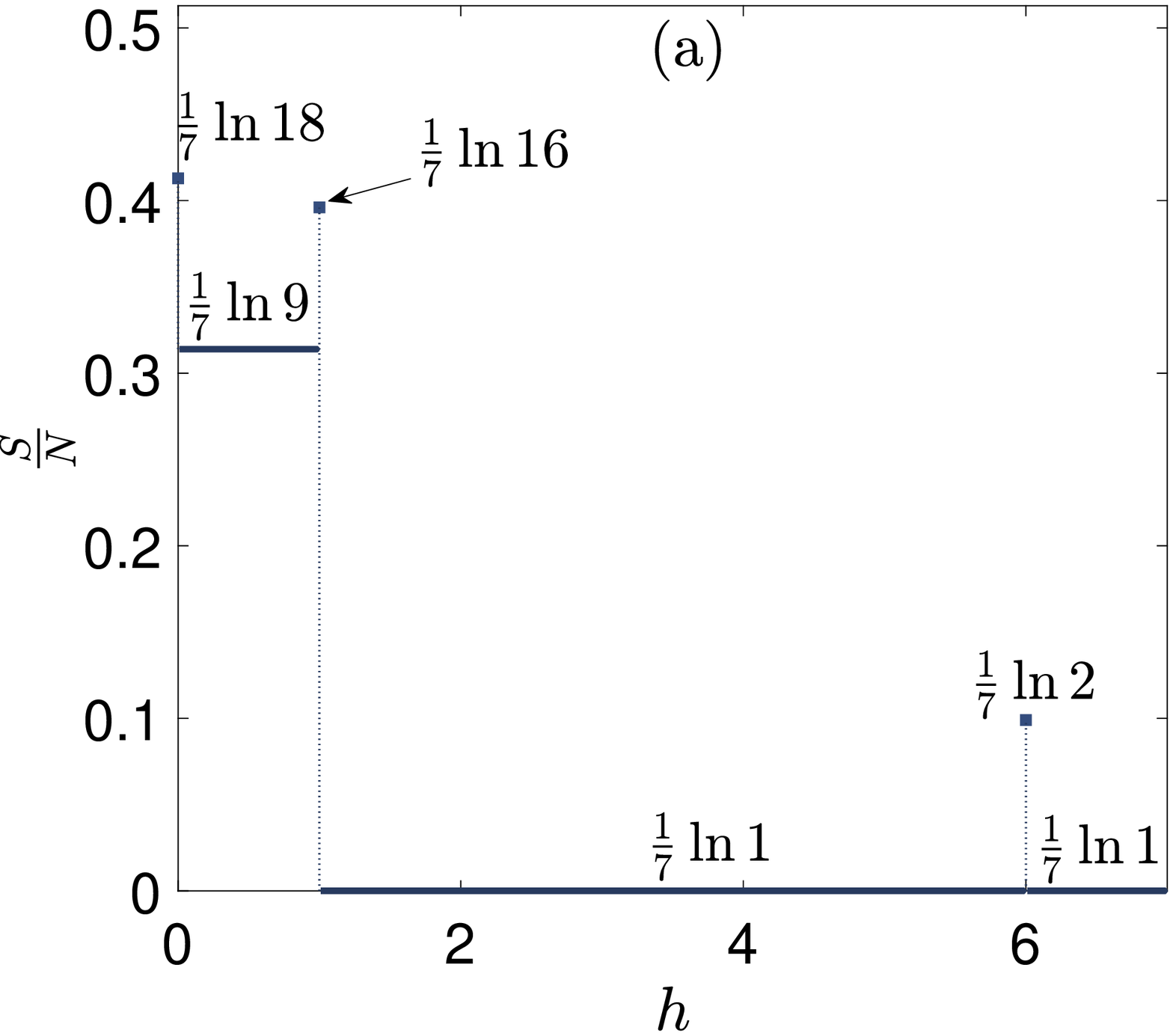}\label{fig:ent_gs_2cs}}  
\subfigure{\includegraphics[scale=0.32,clip]{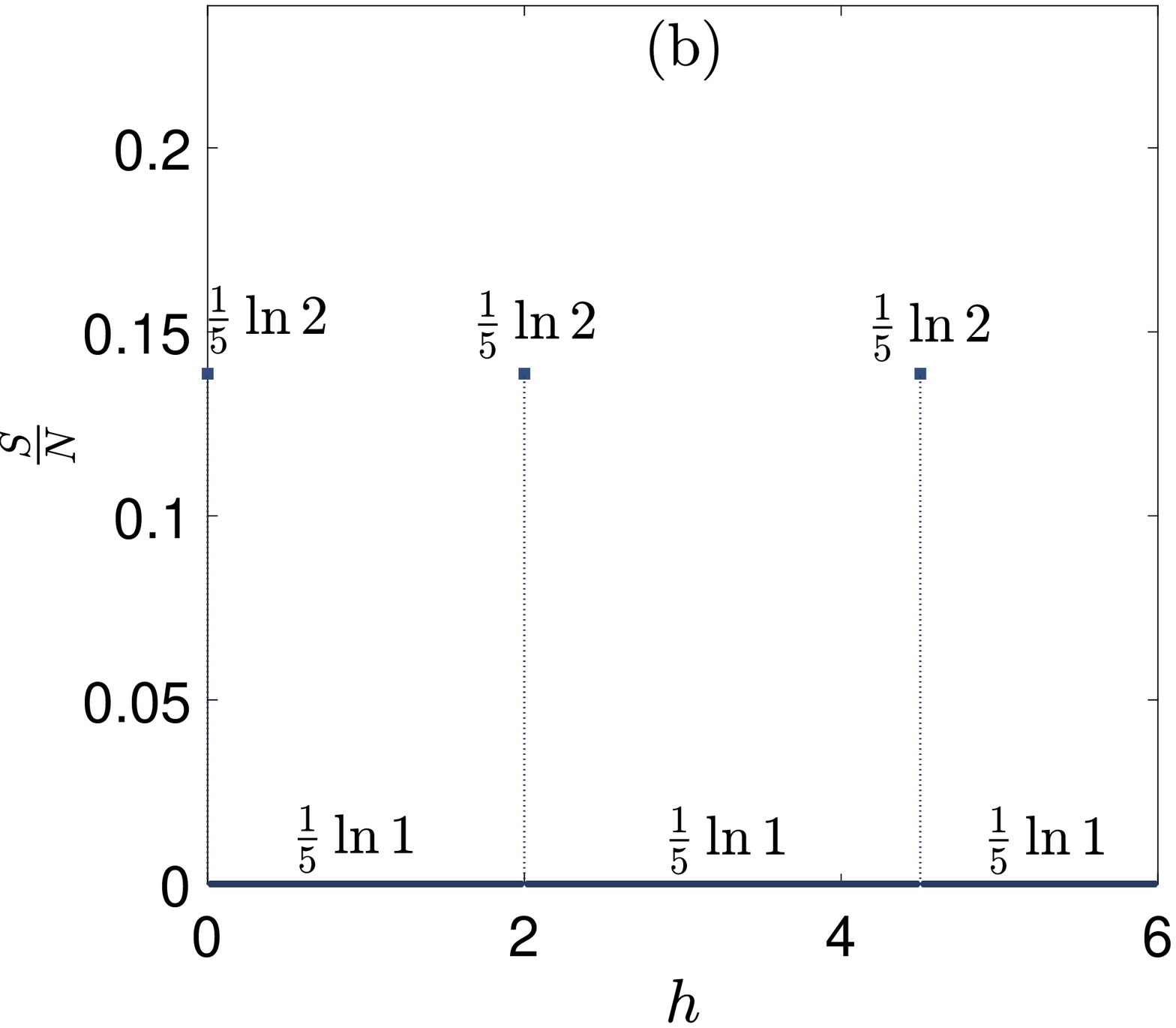}\label{fig:ent_gs_2fs}}\\
\subfigure{\includegraphics[scale=0.32,clip]{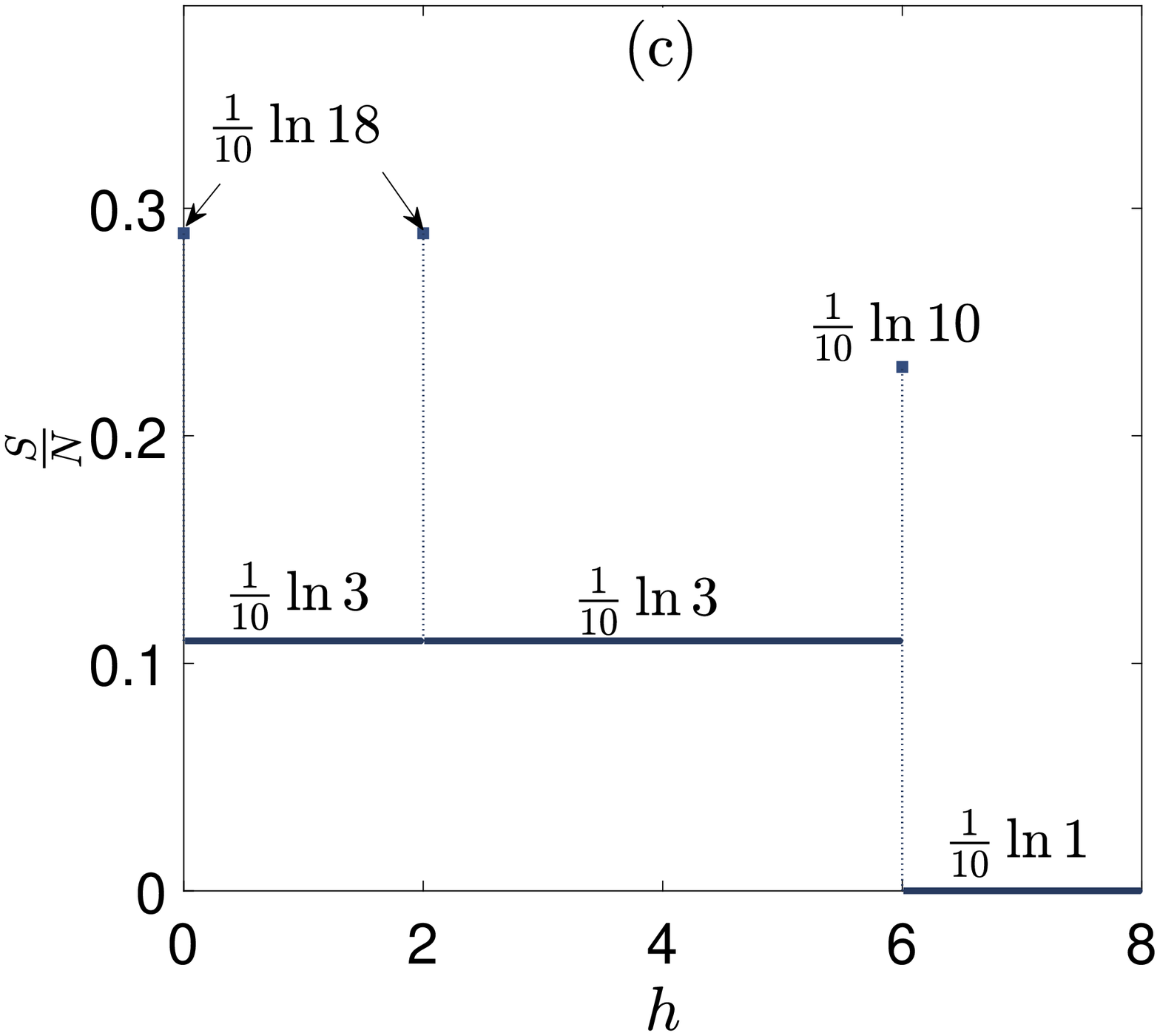}\label{fig:ent_gs_4cs}}
\subfigure{\includegraphics[scale=0.32,clip]{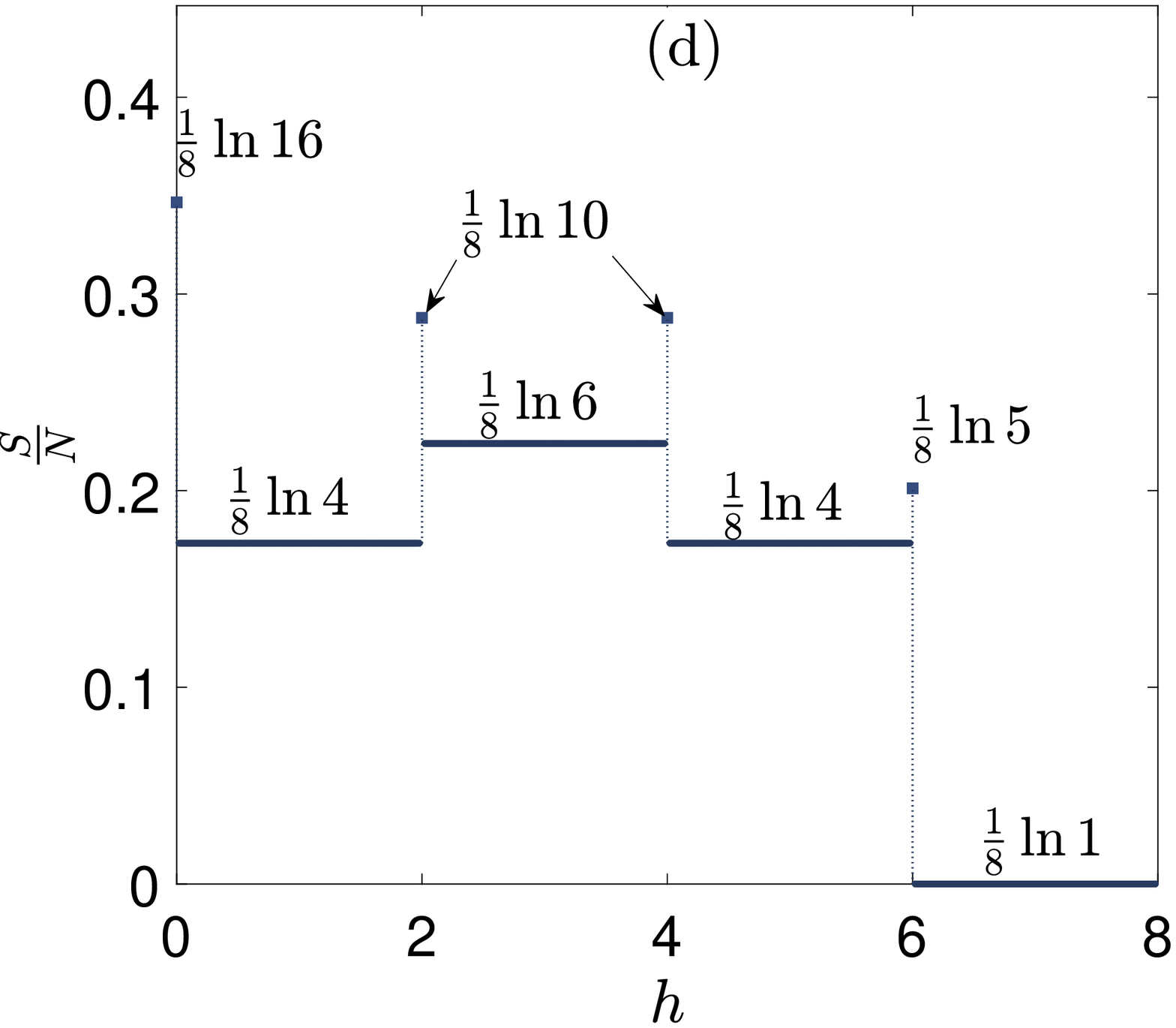}\label{fig:ent_gs_4es}}\\
\subfigure{\includegraphics[scale=0.32,clip]{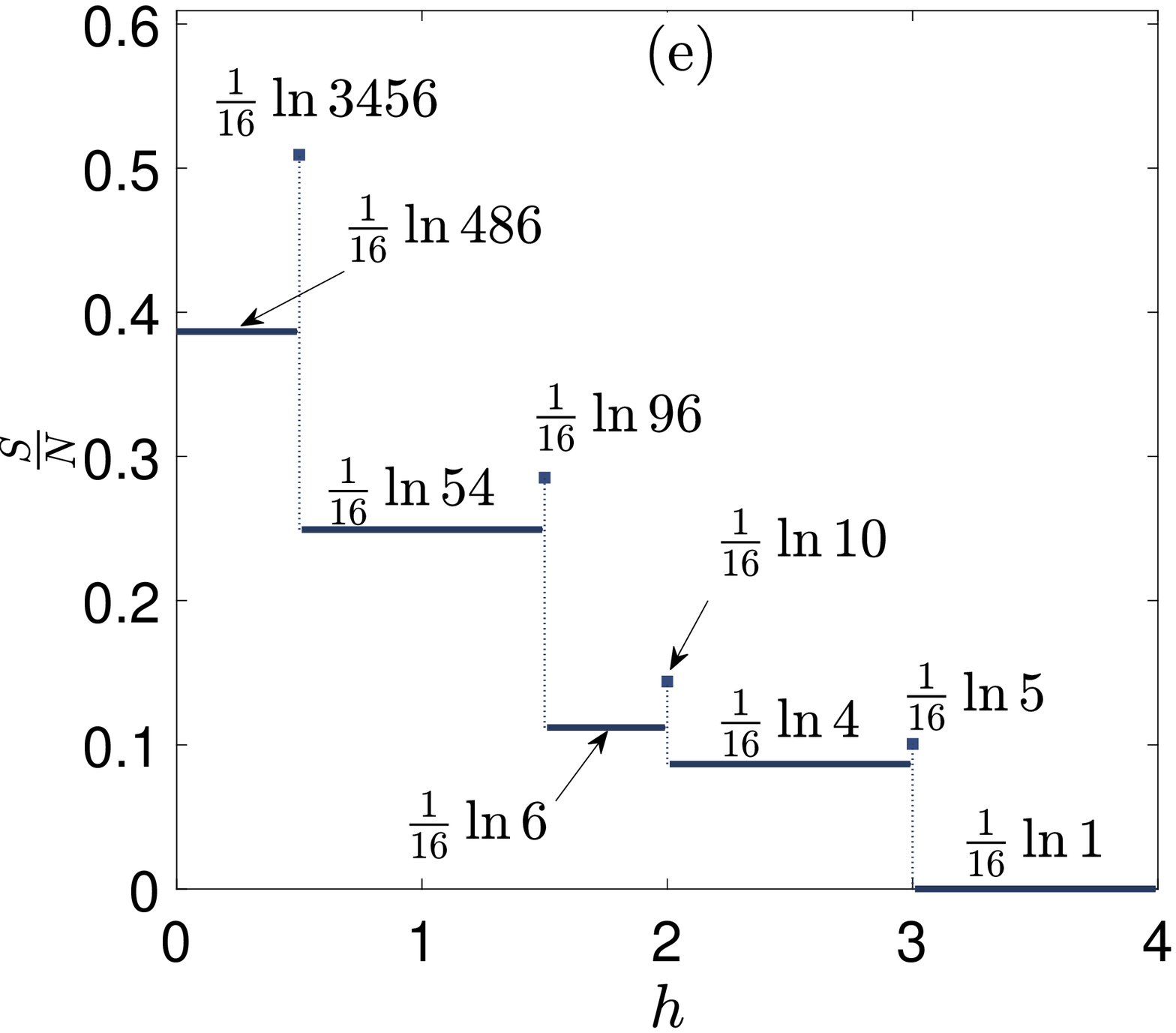}\label{fig:ent_gs_5cs}}
\subfigure{\includegraphics[scale=0.32,clip]{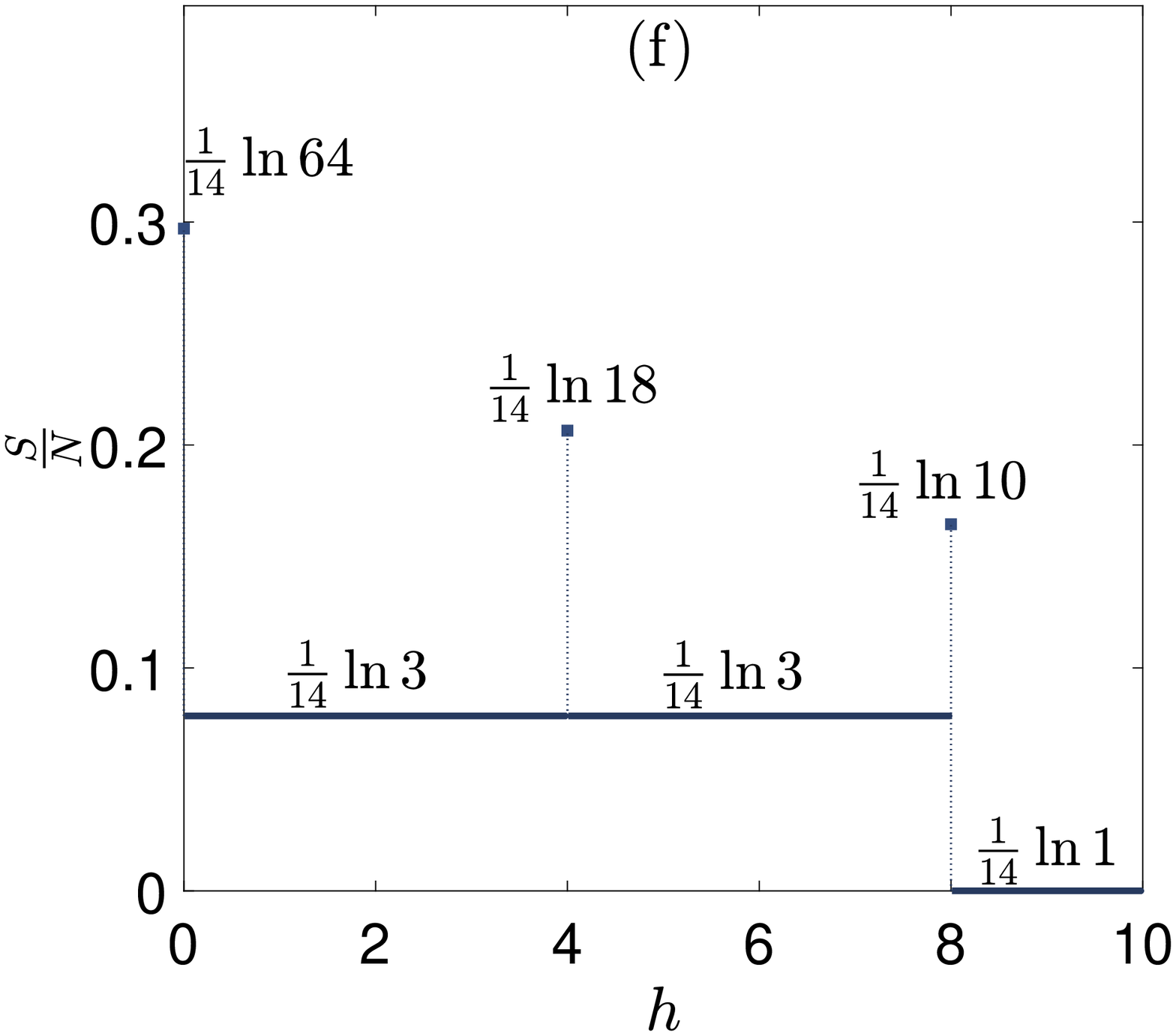}\label{fig:ent_gs_8es}} 
\caption{Field-dependence of the GS entropy density in (a) 2CS, (b) 2FS, (c) 4CS, (d) 4ES, (e) 5CS, and (f) 8ES clusters.} 
\label{fig:ent_gs}
\end{figure}

Zero-temperature magnetization processes in the individual clusters, normalized with respect to the saturation value $m_{sat}=1/2$, are presented in Fig.~\ref{fig:mag_gs}. Let us recall that the GS magnetization process of a single tetrahedron consists of two magnetization plateaux: first one extending within $0 \leq h < 1$ with the height $m/m_{sat}=0$, followed by the second one within $1 < h < 3$ with the height $m/m_{sat}=1/2$, before the saturation phase extending beyond $h_{sat}=3$~\cite{stre15}. Thus, there are two critical fields, $h_{c1}=1$ and $h_{c2}\equiv h_{sat}=3$, at which the magnetization jumps between the respective plateaux.\\
\hspace*{5mm} Fig.~\ref{fig:mag_gs} shows that this scenario changes in the considered clusters, depending on their size and topology. Namely, some clusters preserve the two intermediate plateaux structure but they show quantitative changes in terms of the critical fields and the plateaux heights. In particular, the following clusters show two intermediate plateaux: 2CS with $m/m_{sat}=1/7$ for $h \in (0,1)$ and $m/m_{sat}=5/7$ for $h \in (1,6)$, 2FS with $m/m_{sat}=1/5$ for $h \in (0,2)$ and $m/m_{sat}=3/5$ for $h \in (2,9/2)$, 4CS with $m/m_{sat}=1/5$ for $h \in (0,2)$ and $m/m_{sat}=3/5$ for $h \in (2,6)$, and 8ES with $m/m_{sat}=3/7$ for $h \in (0,4)$ and $m/m_{sat}=5/7$ for $h \in (4,8)$. It is worthwhile noting that in the clusters with no zero-magnetization plateau the lowest critical field occurs at $h_{c1}=0$ and, therefore, in the above mentioned cases there are three critical fields. In the cases of the 4ES and 5CS clusters, there are even four critical fields with three and four plateaux, respectively. Namely, in the 4ES case three plateaux of the heights $m/m_{sat}=1/4,1/2$, and $3/4$ lie between the critical fields $h_{c1}=0,h_{c2}=2,h_{c3}=4$ and $h_{c4}=6$ and in the 5CS case four plateaux of the heights $m/m_{sat}=0,1/2,3/4$ and $7/8$ are separated by the critical fields $h_{c1}=1/2,h_{c2}=3/2,h_{c3}=2$ and $h_{c4}=3$. We also note that the magnetization values in the respective critical fields generally differ from the values inside the plateaux (see Fig.~\ref{fig:mag_gs}), due to the coexistence of states with different degeneracies, as will be discussed below.\\
\hspace*{5mm} The corresponding entropy densities are plotted in Fig.~\ref{fig:ent_gs}. The values pertaining to the respective plateaux as well as those at the critical fields are presented in the form $S/N=(1/N)\ln W$ to explicitly provide degeneracies in the respective phases. A general tendency of an increasing external field is to gradually eliminate the number of degenerate states ultimately to a single one corresponding to the fully polarized phase above the saturation field value $h_{sat}$. Fig.~\ref{fig:ent_gs} shows a variety of ways how the GS degeneracy is gradually lifted in different clusters. The cluster 2FS is an example of a system with low GS degeneracy (two-fold), which is lifted already at infinitesimally small field values. On the other hand, the clusters 4CS, 4ES, 5CS and 8ES require the fields larger than the saturation value to completely remove the degeneracy. The 4ES cluster is the only case in which an increasing field even enhances the degeneracy within $2 < h < 4$ before it is eliminated at larger fields. Such a non-monotonous behavior is a consequence of the frustration and it has already been observed in some other frustrated Ising nanoclusters~\cite{zuko15,zuko18}. \\
\hspace*{5mm} As already mentioned above, right in the critical field values degeneracy is locally increased due to contributions from different coexisting states, which is reflected in spikes in the entropy density and isolated points in the magnetization plateaux in field-dependencies of the respective quantities. We note that similar behavior with isolated values of the magnetization and entropy at the field-induced first-order transitions in the ground state have also been reported in some other antiferromagnetic Ising systems~\cite{metc78} including the tetrahedron recursive lattice~\cite{jurc14,jurc17}.

\subsubsection{Finite temperatures}

\begin{figure}[t!]   
\centering 
\subfigure{\includegraphics[scale=0.32,clip]{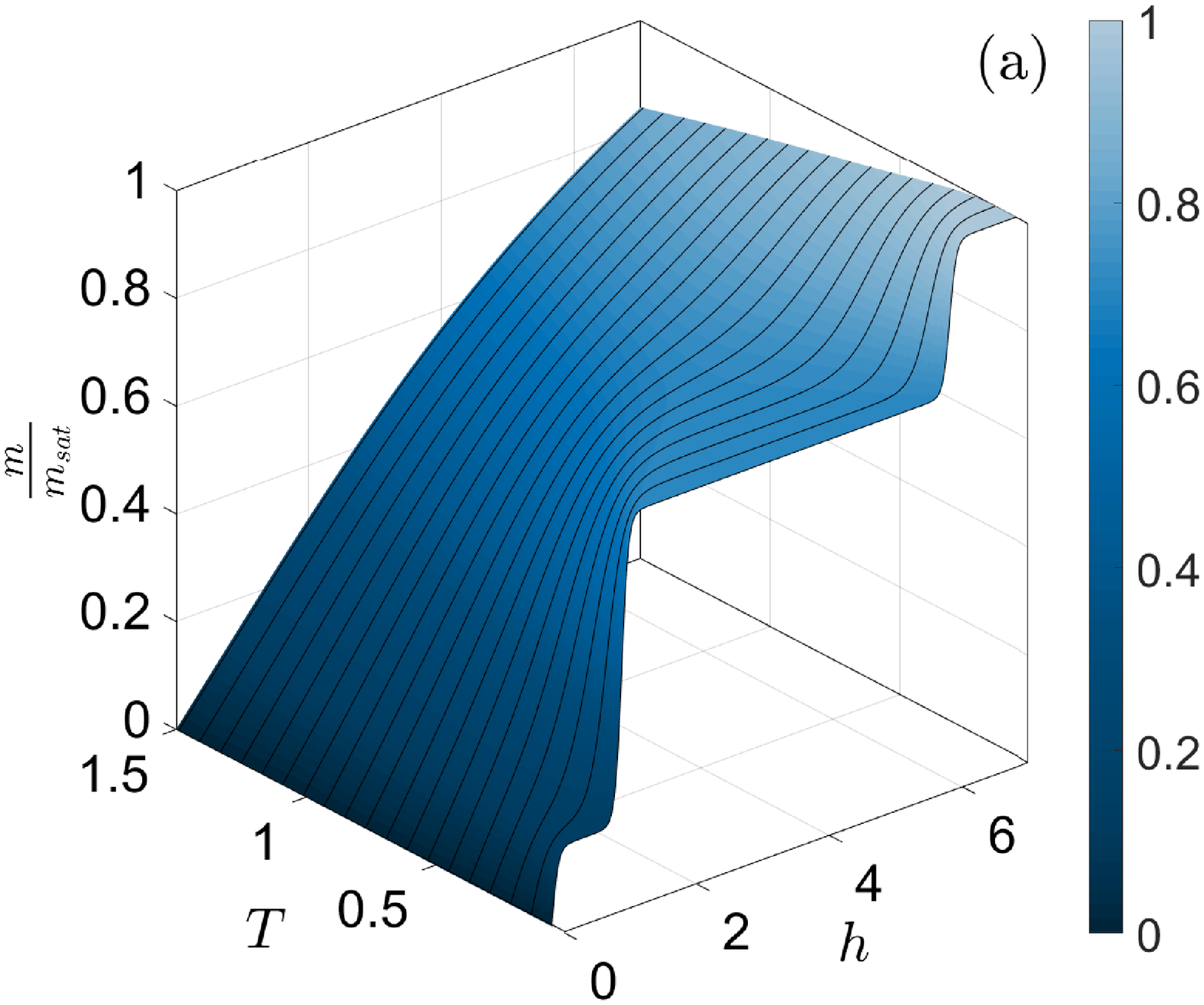}\label{fig:mag_side_2cs}} 
\subfigure{\includegraphics[scale=0.32,clip]{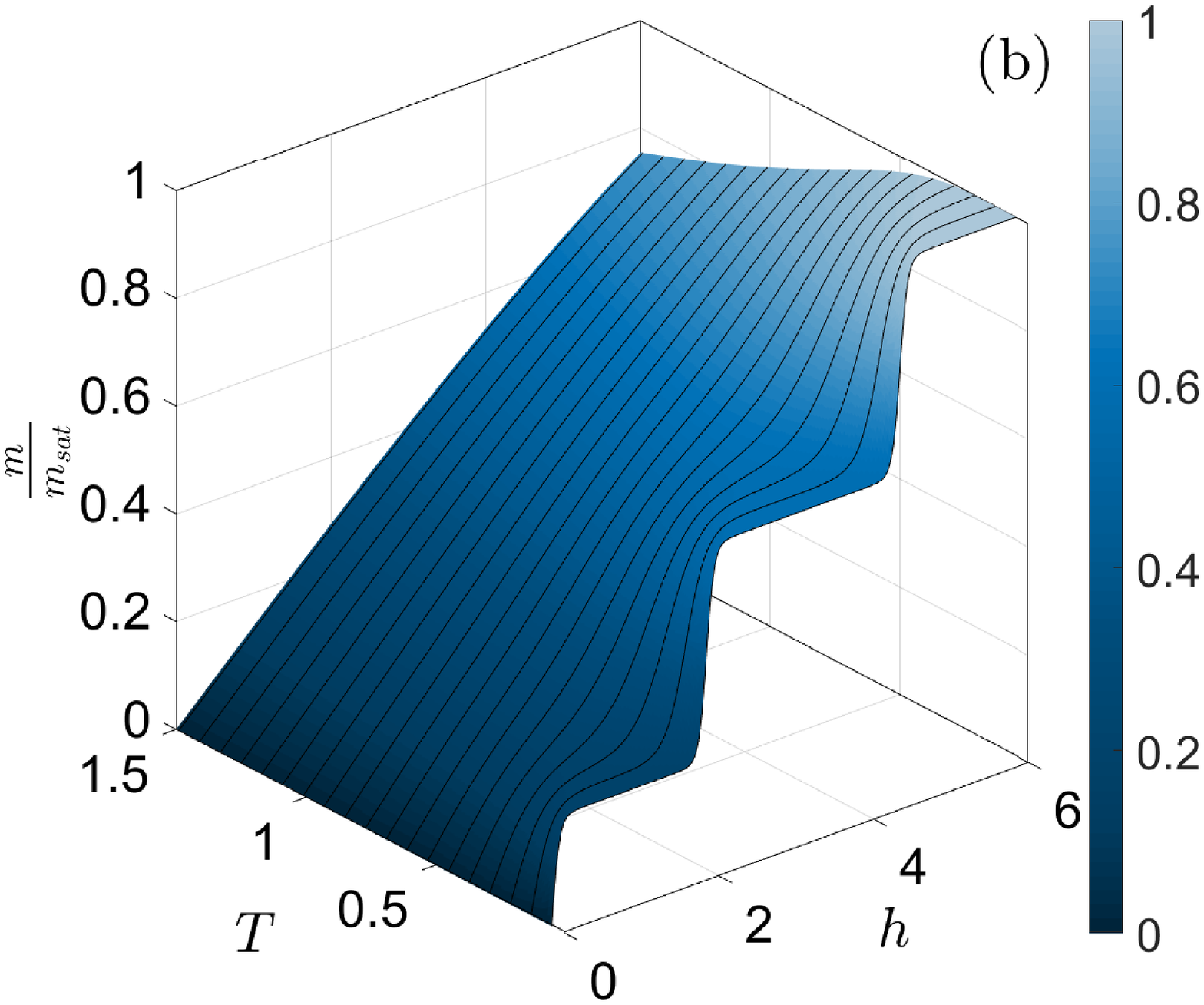}\label{fig:mag_side_2fs}}\\
\subfigure{\includegraphics[scale=0.32,clip]{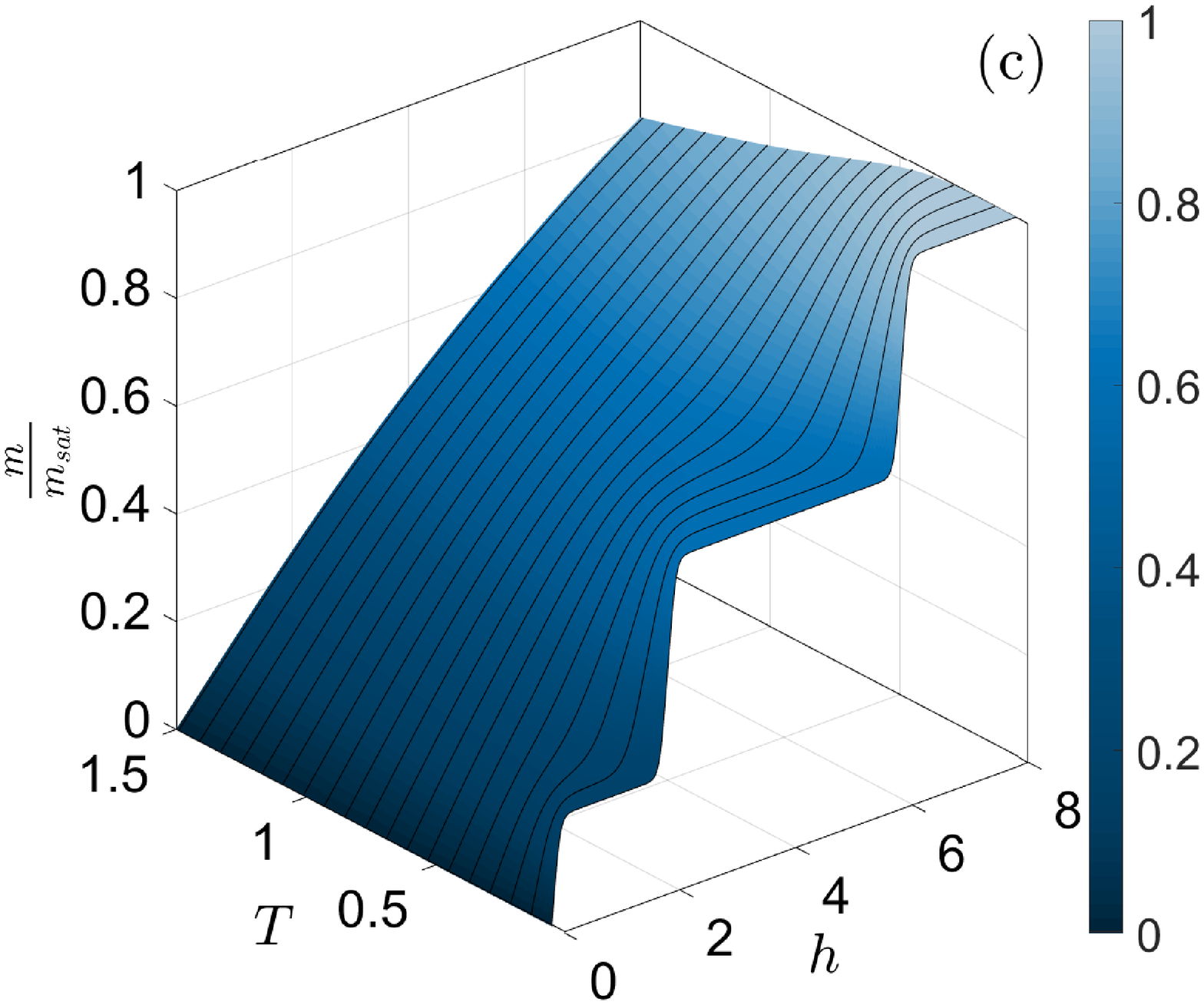}\label{fig:mag_side_4cs}}
\subfigure{\includegraphics[scale=0.32,clip]{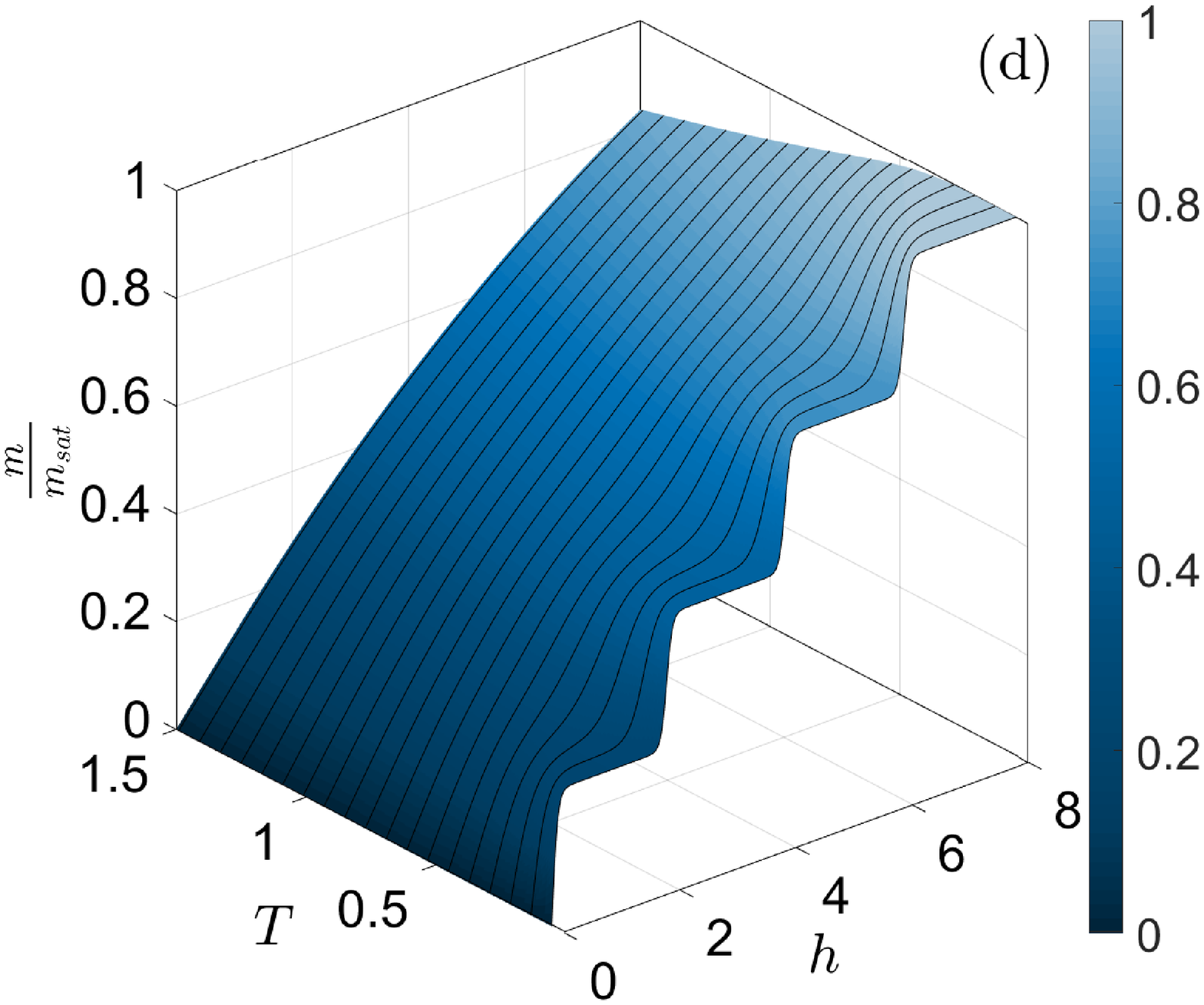}\label{fig:mag_side_4es}}\\
\subfigure{\includegraphics[scale=0.32,clip]{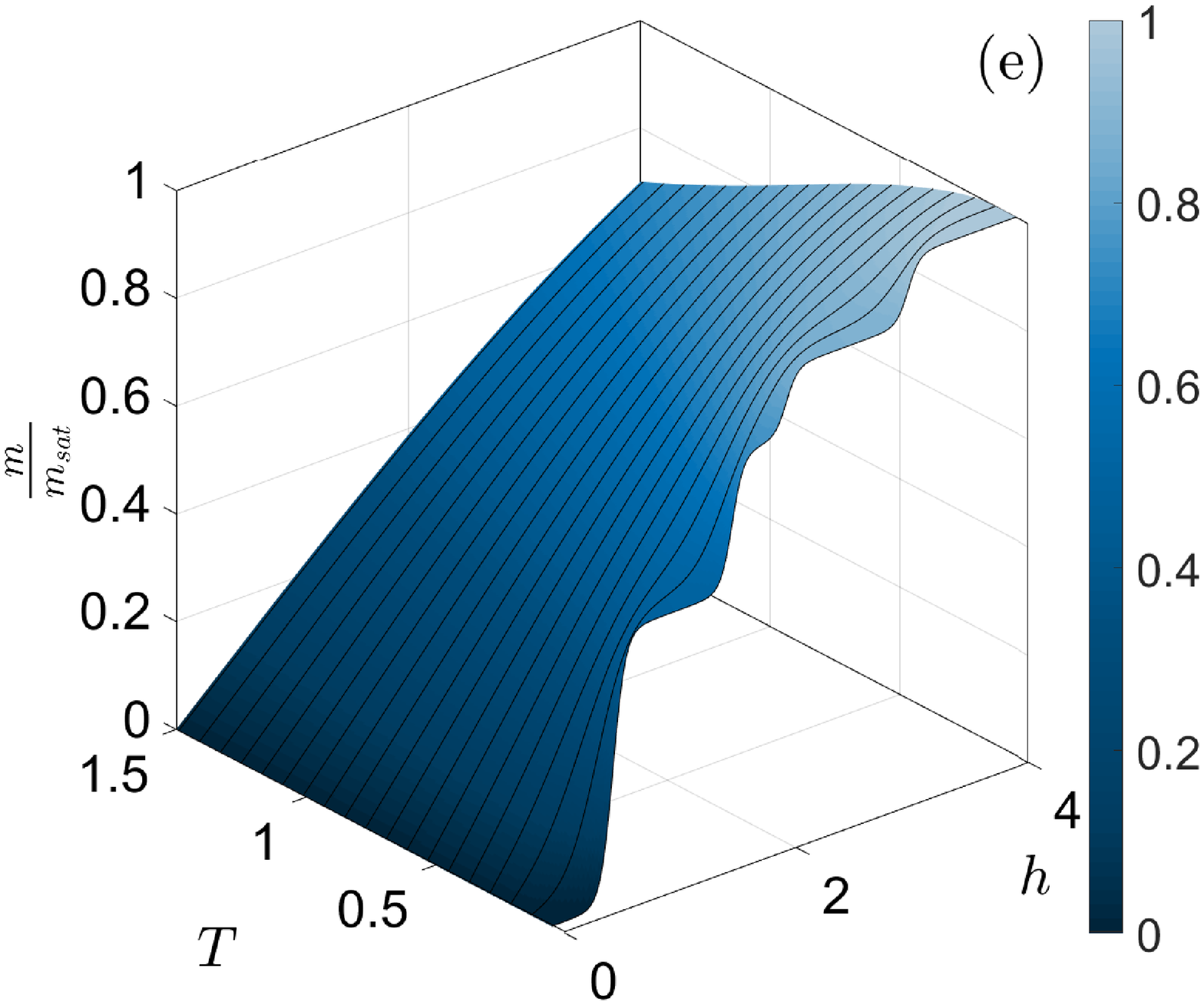}\label{fig:mag_side_5cs}}
\subfigure{\includegraphics[scale=0.32,clip]{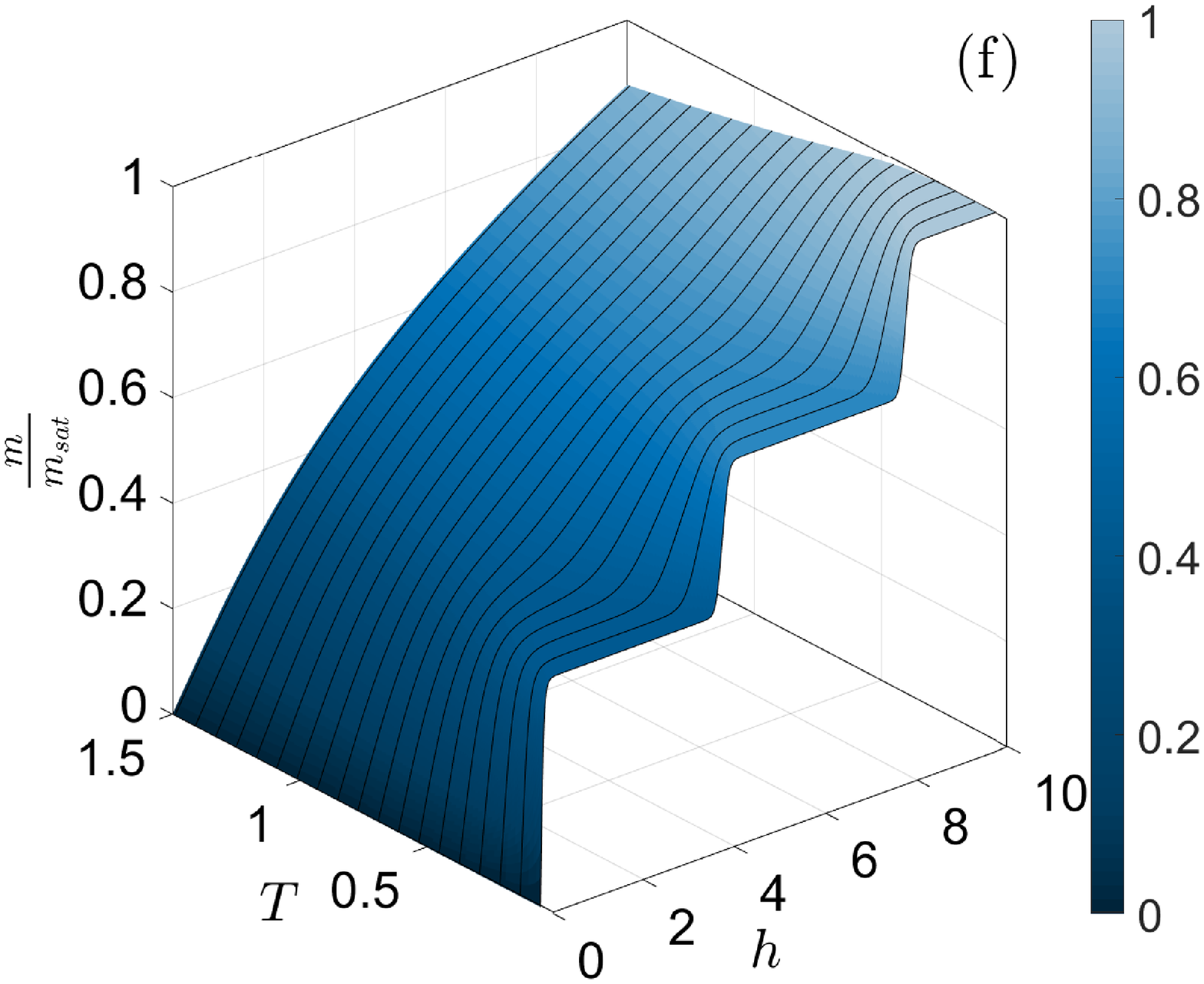}\label{fig:mag_side_8es}} 
\caption{Magnetization in $T$-$h$ plane, for (a) 2CS, (b) 2FS, (c) 4CS, (d) 4ES, (e) 5CS, and (f) 8ES clusters.} 
\label{fig:mag_fin}
\end{figure}

\begin{figure}[t!]   
\centering 
\subfigure{\includegraphics[scale=0.32,clip]{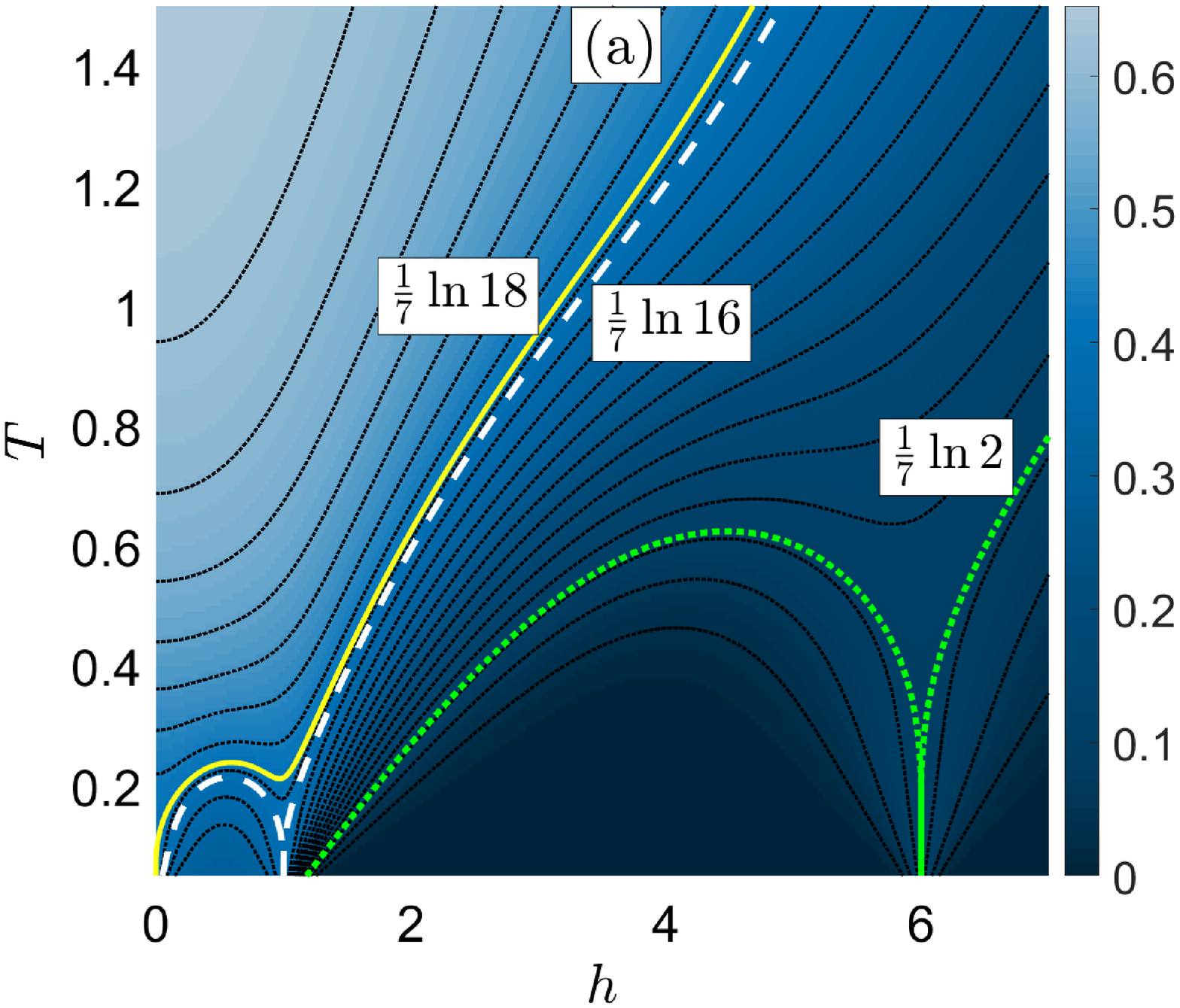}\label{fig:ent_top_2cs}}  
\subfigure{\includegraphics[scale=0.32,clip]{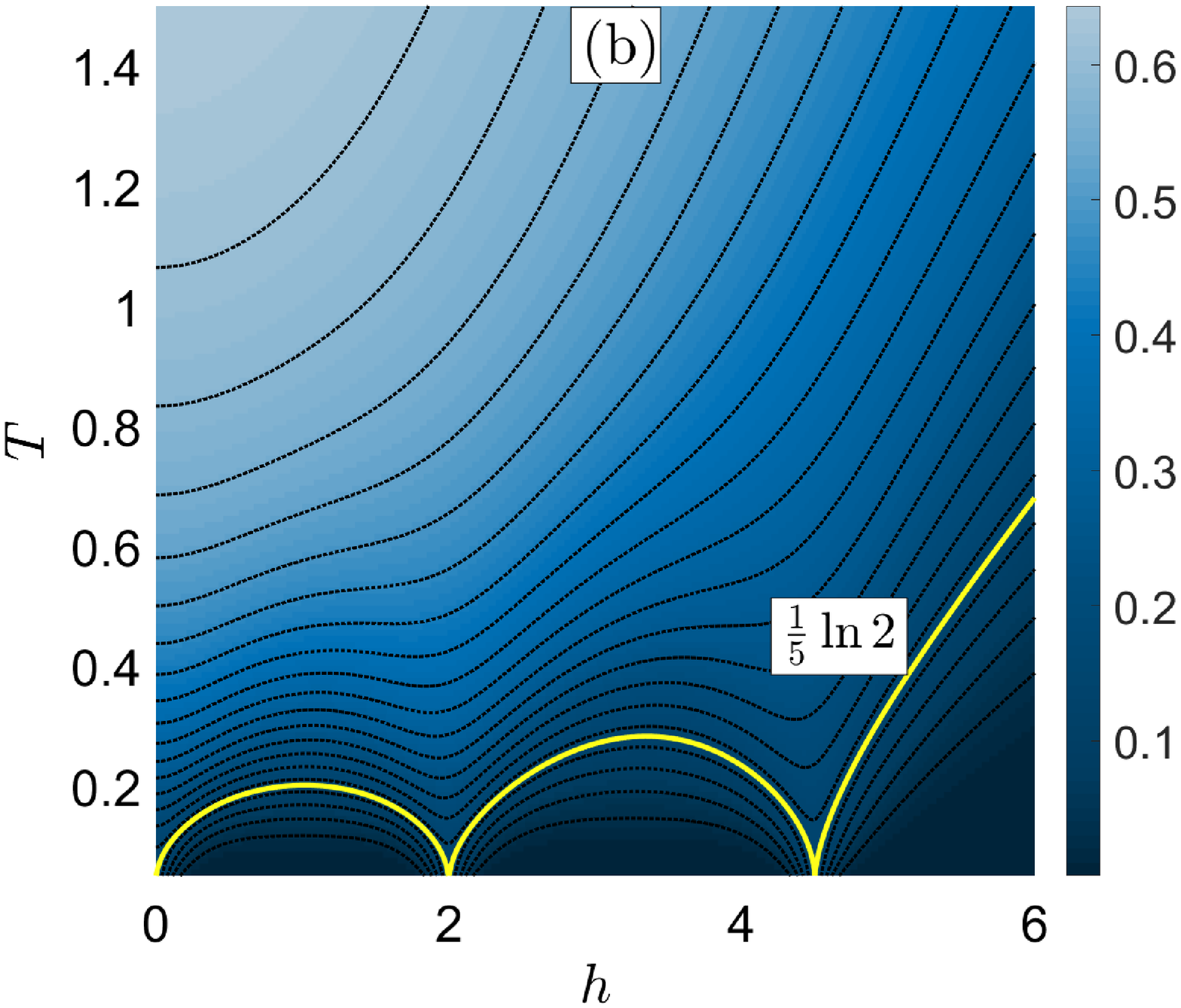}\label{fig:ent_top_2fs}}\\
\subfigure{\includegraphics[scale=0.32,clip]{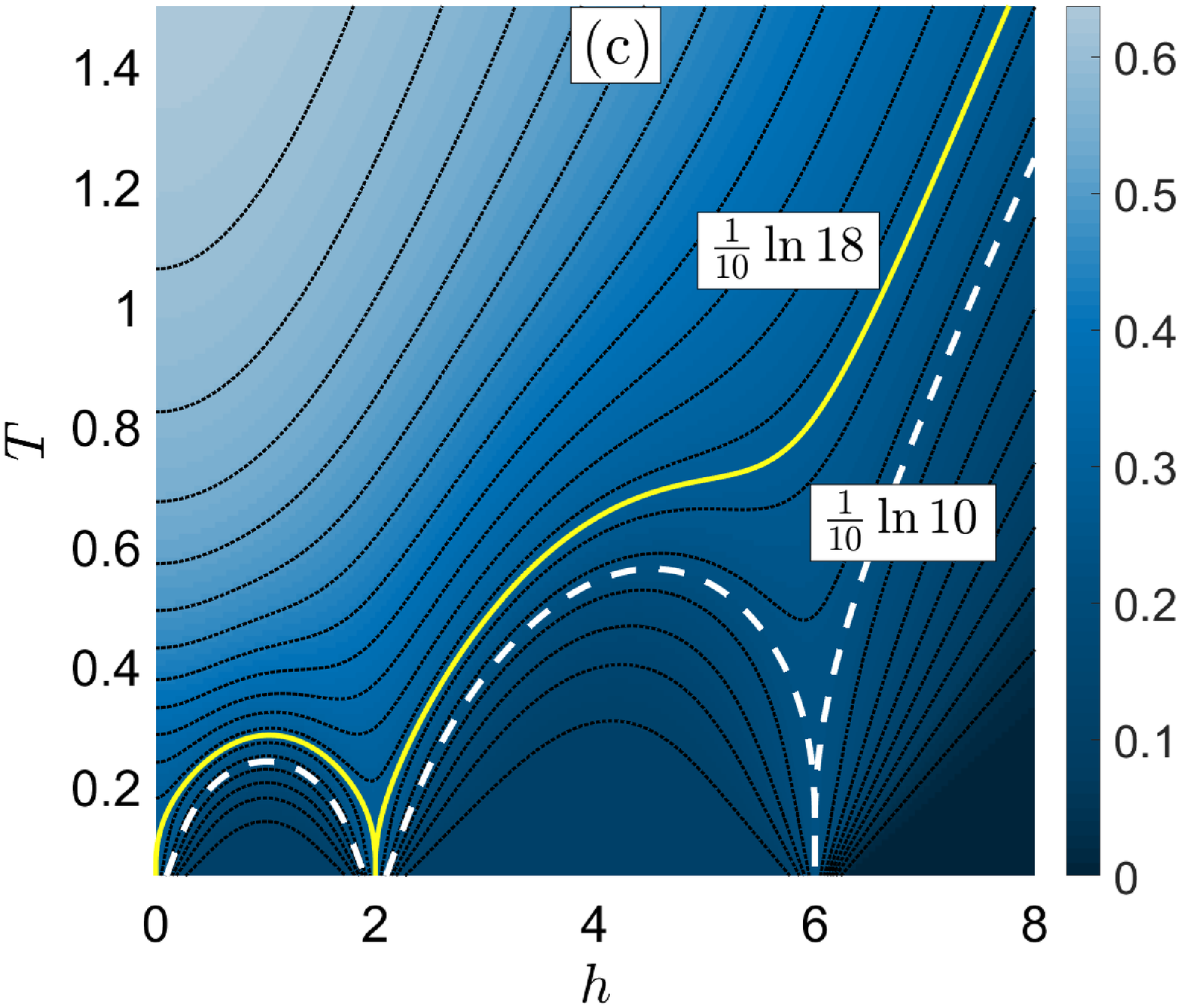}\label{fig:ent_top_4cs}}
\subfigure{\includegraphics[scale=0.32,clip]{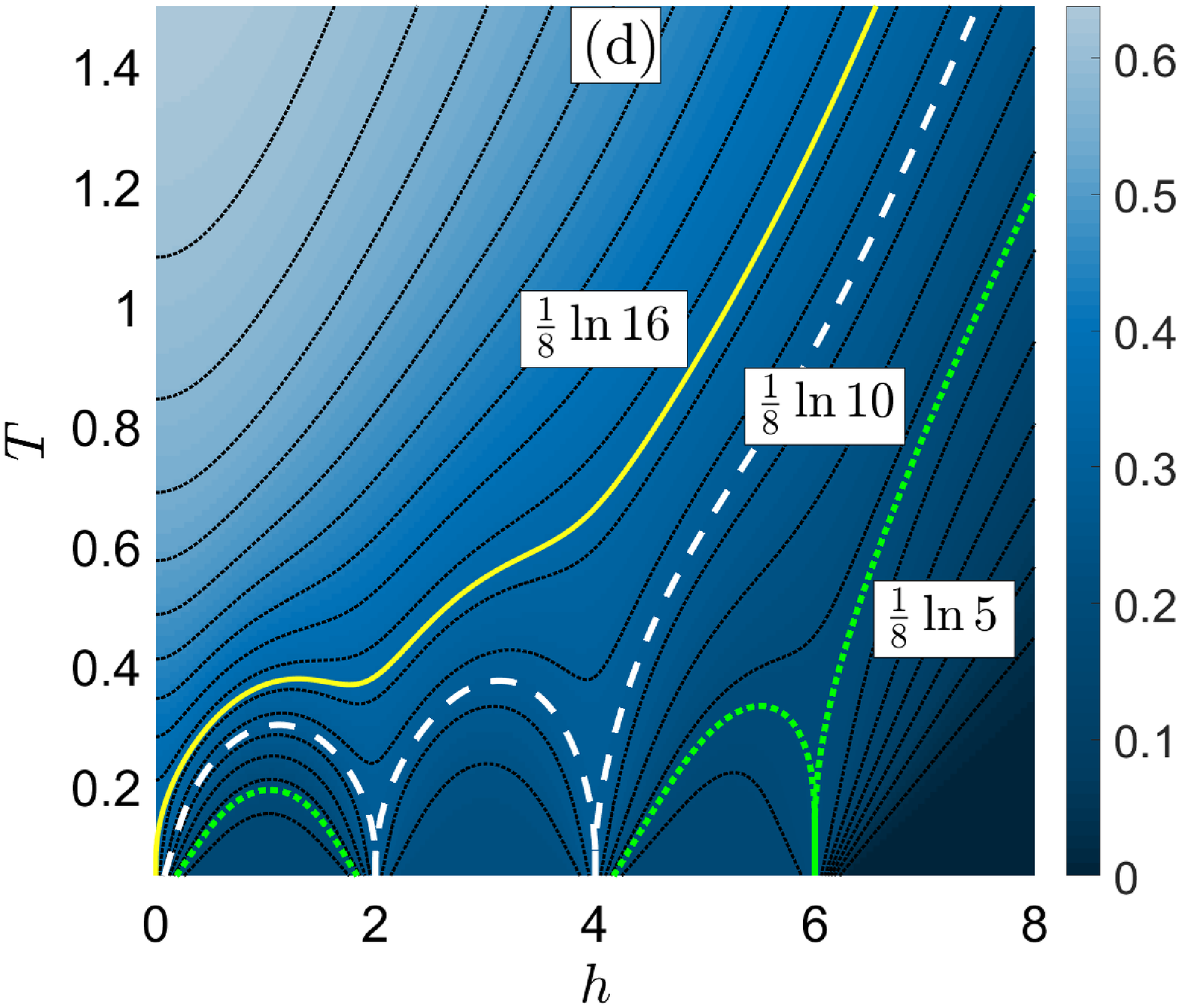}\label{fig:ent_top_4es}}\\
\subfigure{\includegraphics[scale=0.32,clip]{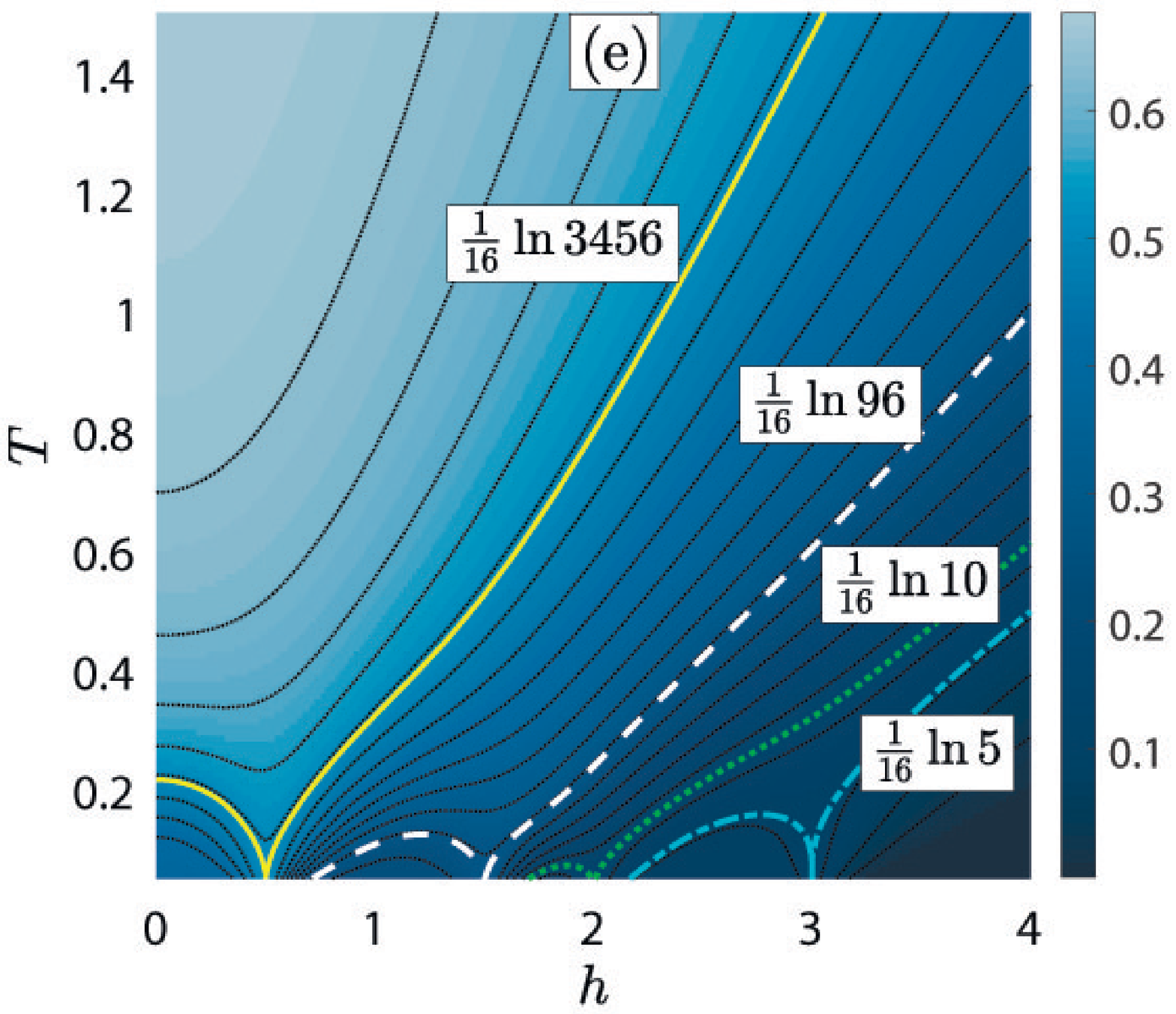}\label{fig:ent_top_5cs}}
\subfigure{\includegraphics[scale=0.32,clip]{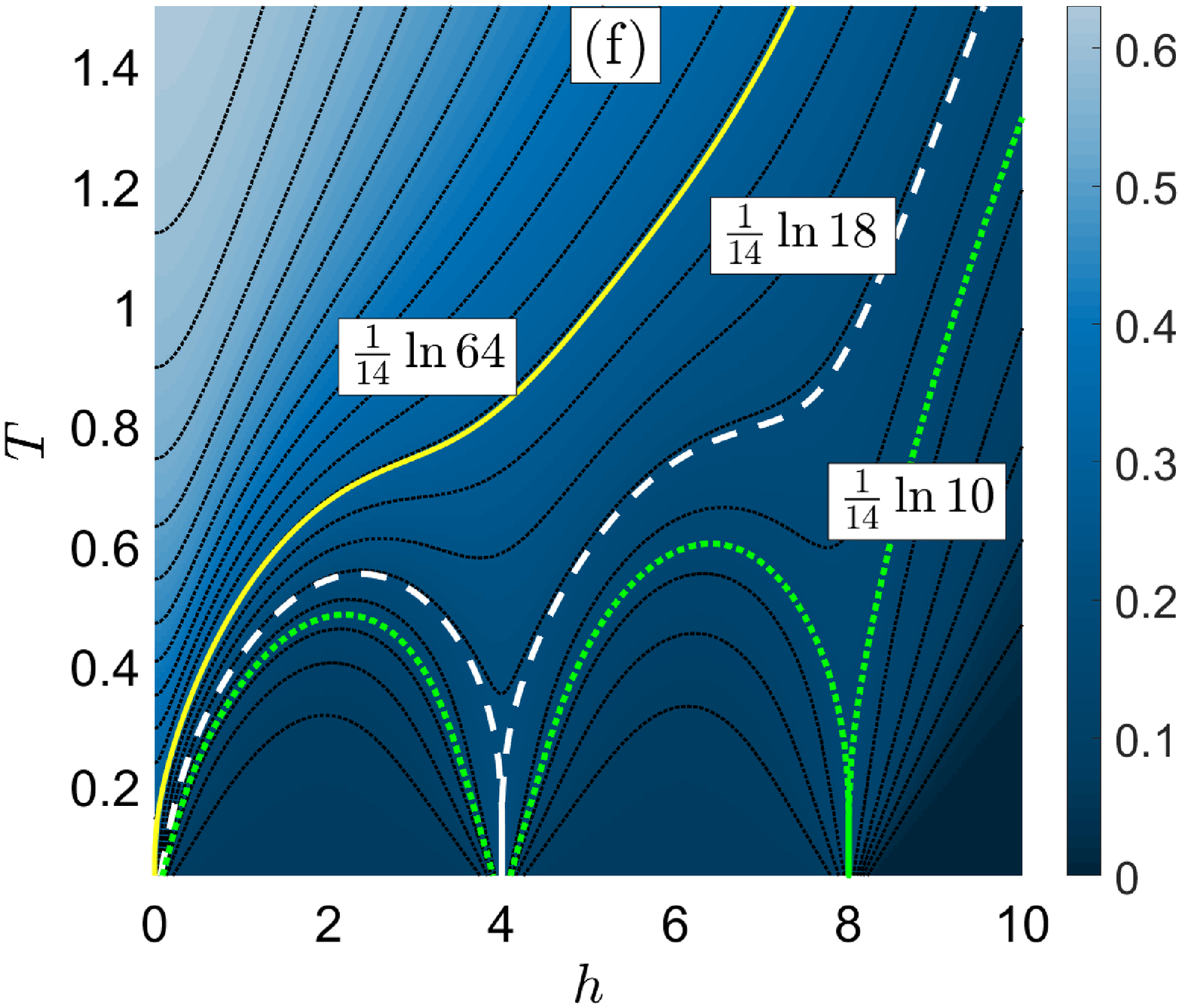}\label{fig:ent_top_8es}} 
\caption{Entropy density in $T$-$h$ plane, for (a) 2CS, (b) 2FS, (c) 4CS, (d) 4ES, (e) 5CS, and (f) 8ES clusters. The highlighted isentropes correspond to the GS residual entropies at the respective critical fields.} 
\label{fig:ent_fin}
\end{figure}

\begin{figure}[t!]   
\centering 
\subfigure{\includegraphics[scale=0.32,clip]{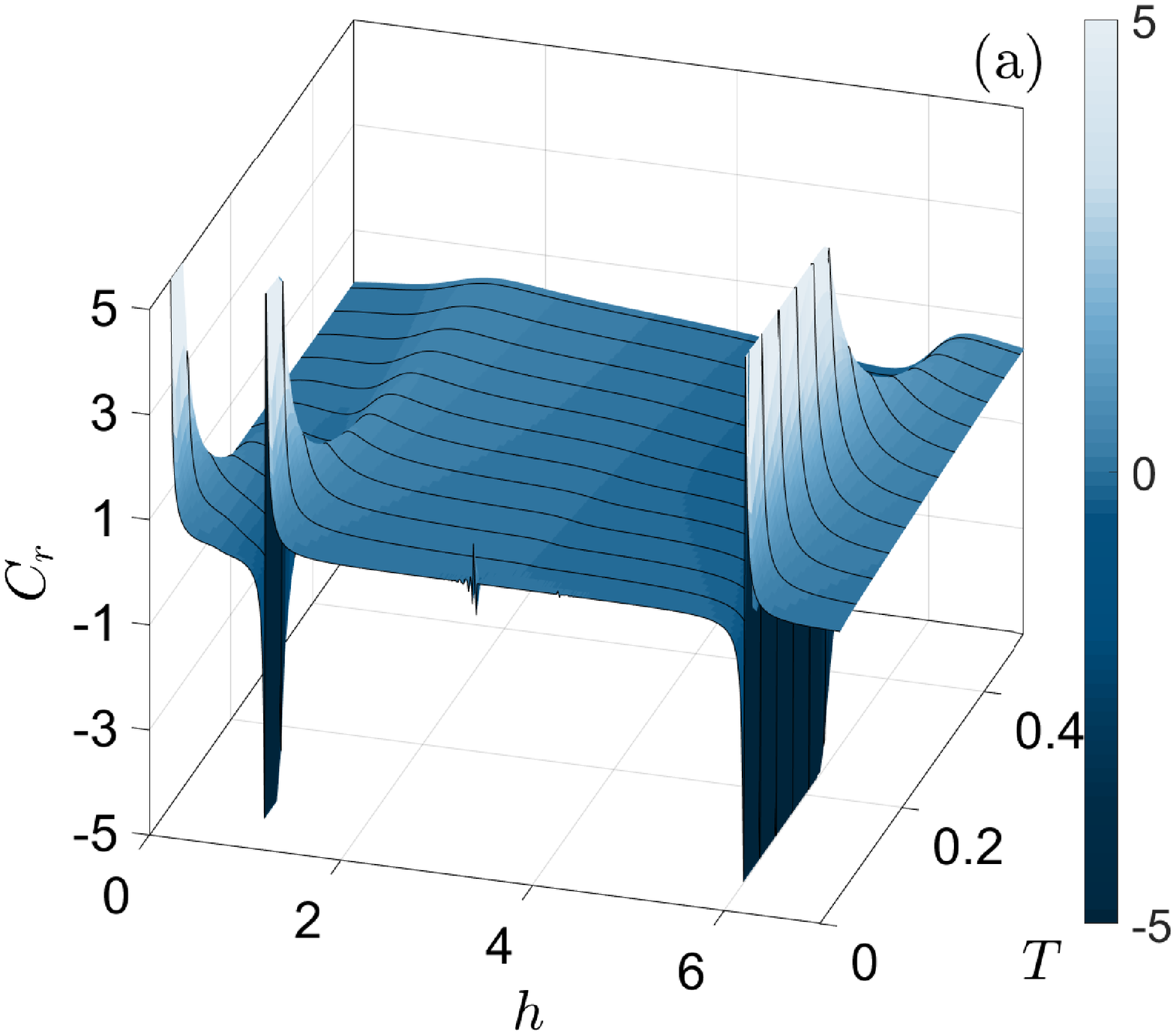}\label{fig:clr_h_2cs}}  
\subfigure{\includegraphics[scale=0.32,clip]{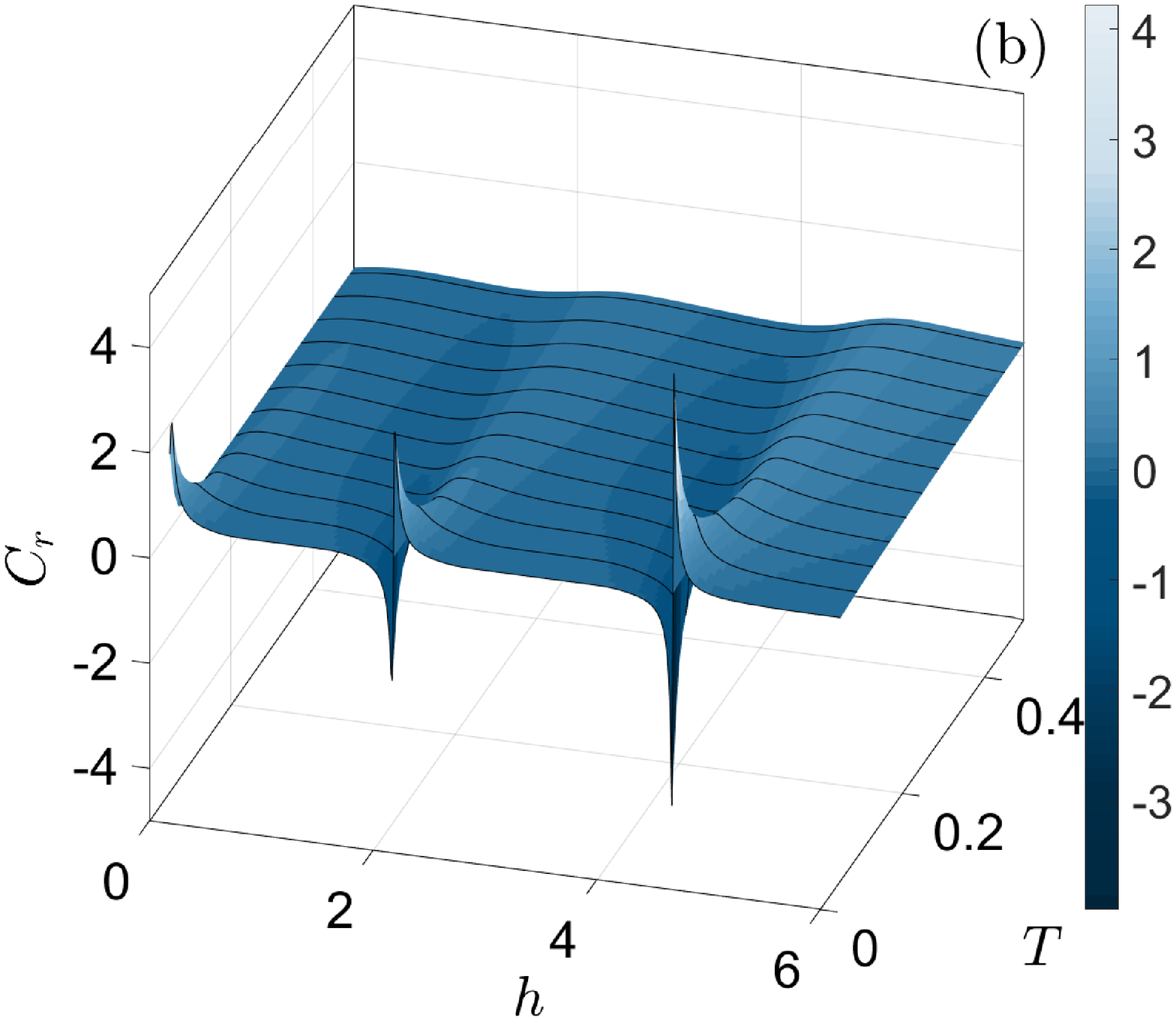}\label{fig:clr_h_2fs}}\\
\subfigure{\includegraphics[scale=0.32,clip]{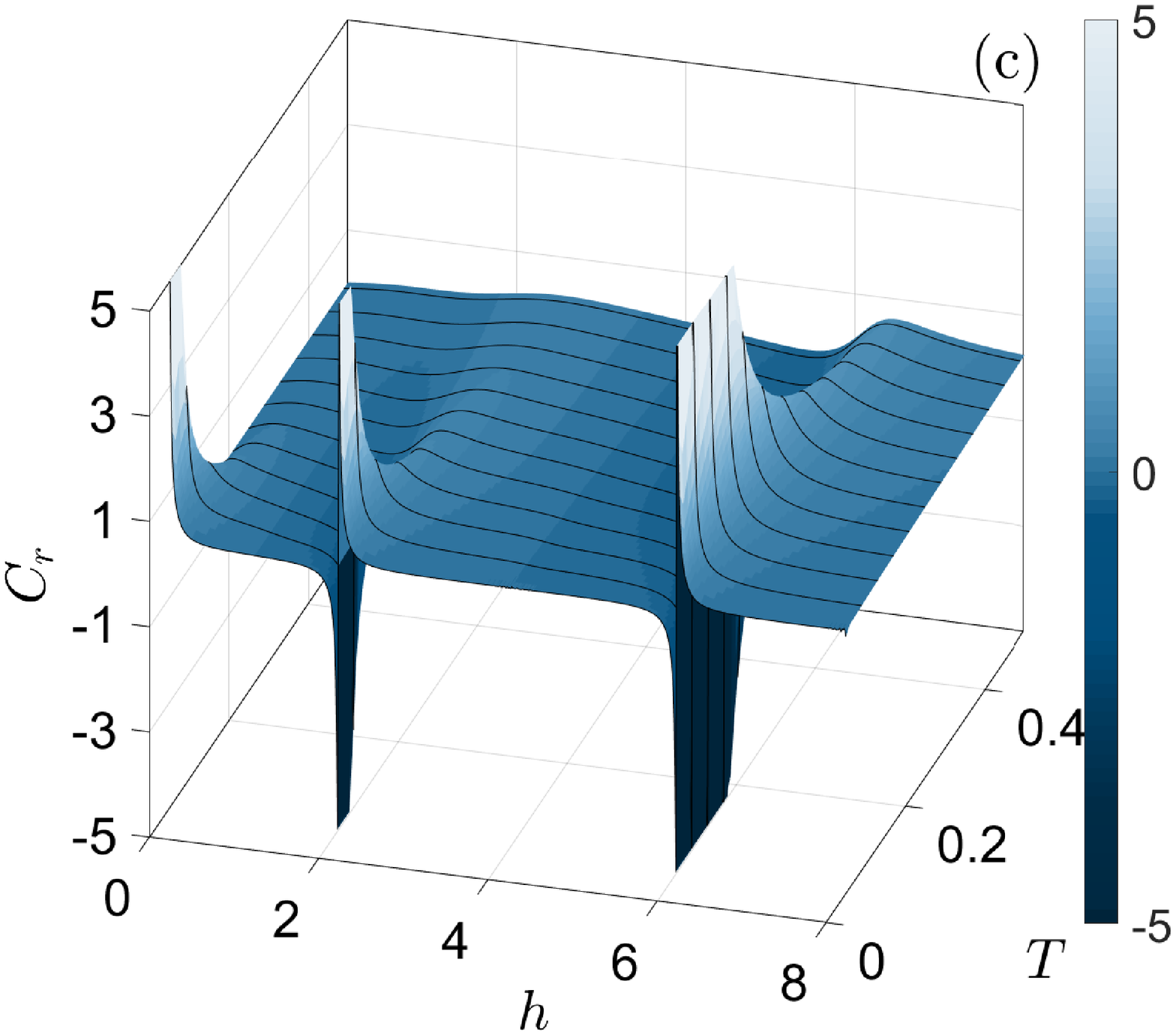}\label{fig:clr_h_4cs}}
\subfigure{\includegraphics[scale=0.32,clip]{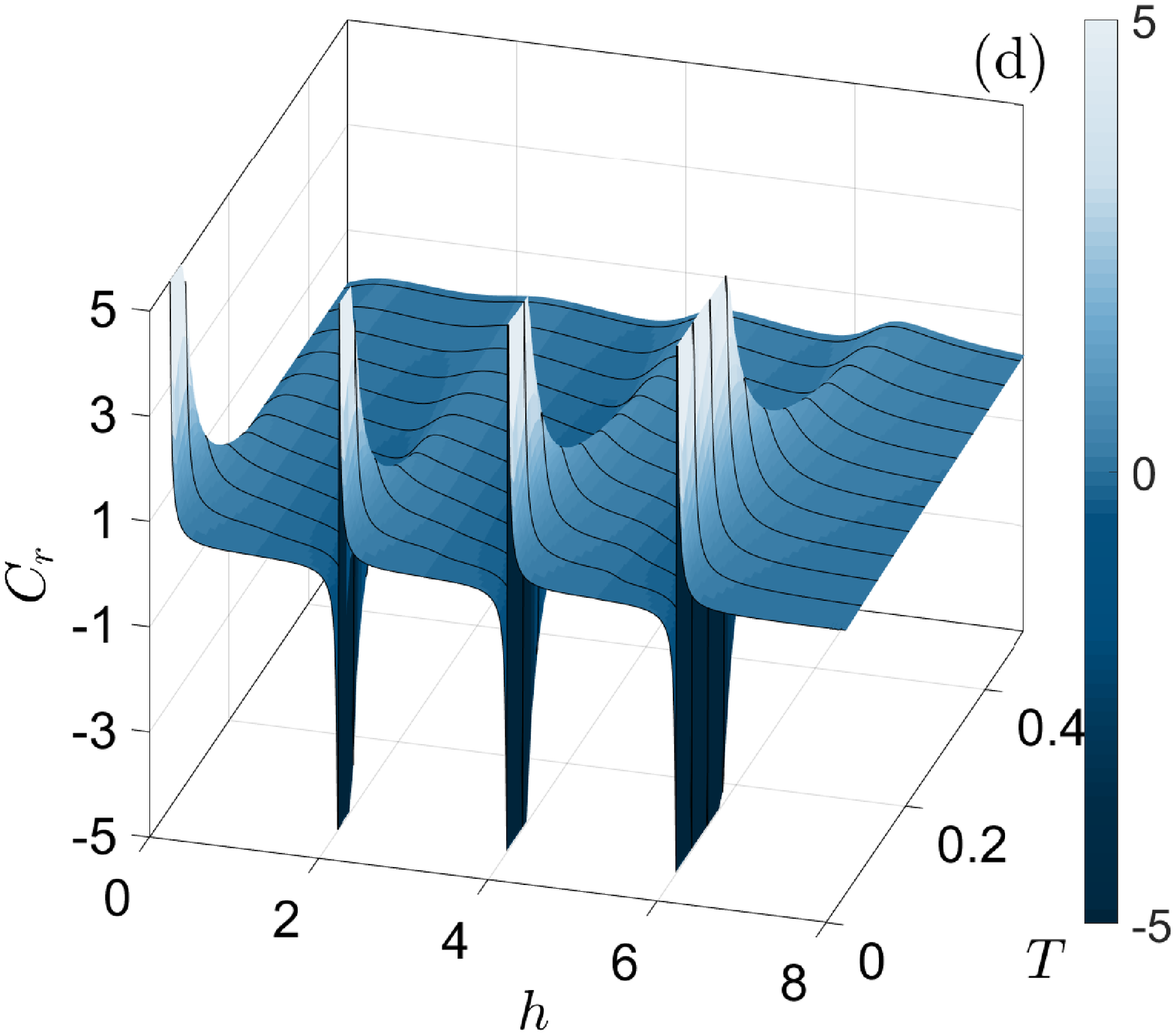}\label{fig:clr_h_4es}}\\
\subfigure{\includegraphics[scale=0.32,clip]{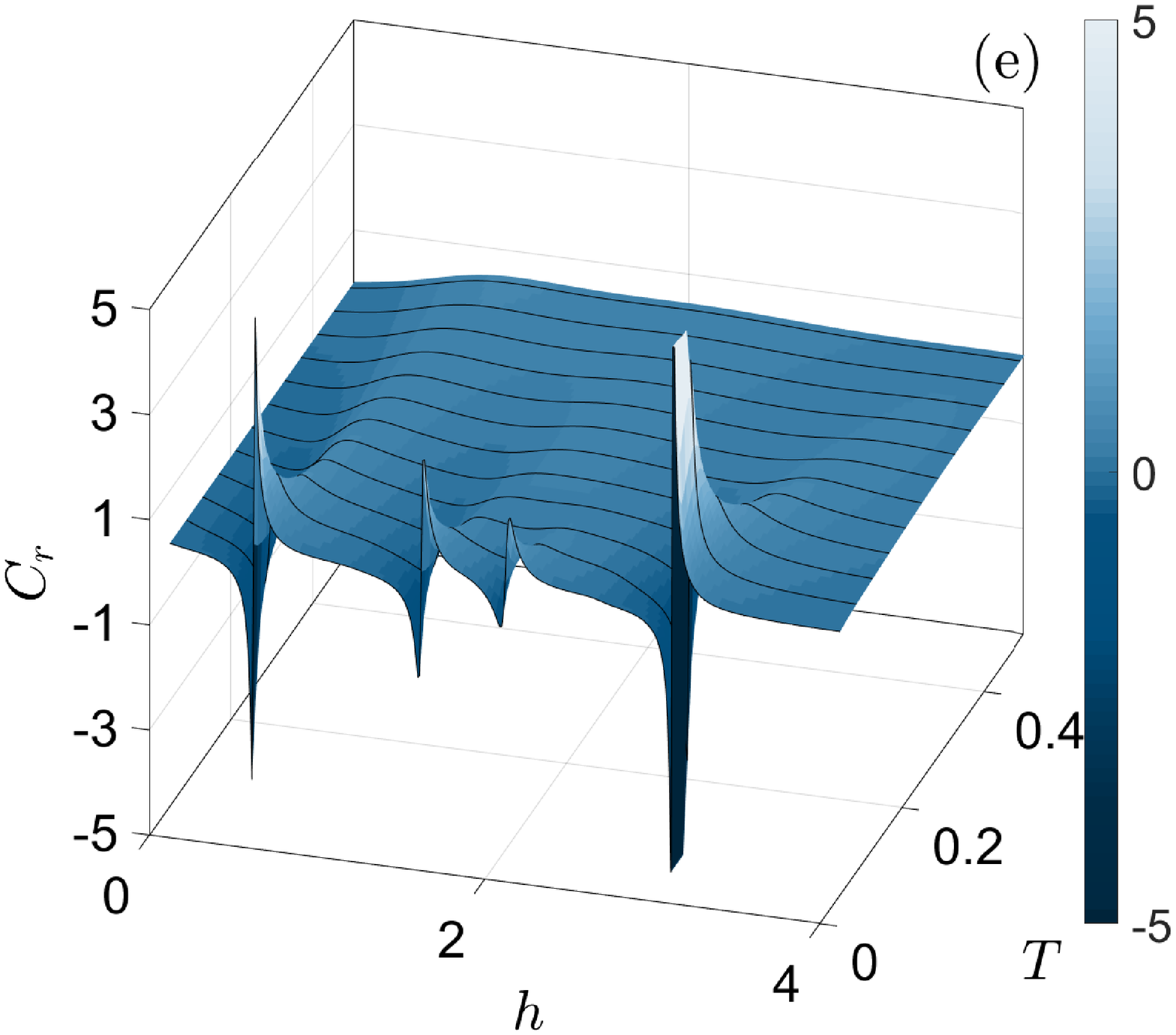}\label{fig:clr_h_5cs}}
\subfigure{\includegraphics[scale=0.32,clip]{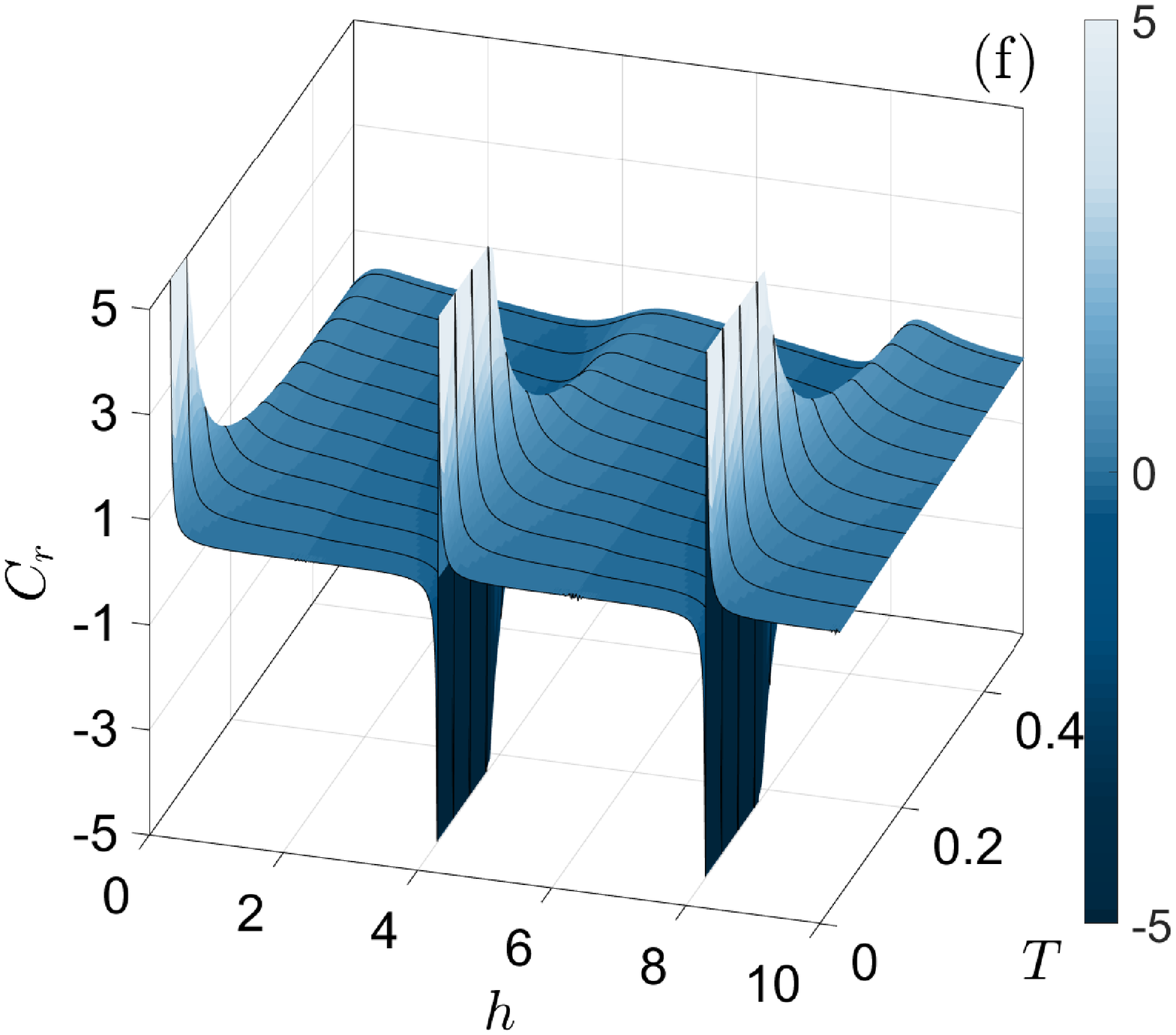}\label{fig:clr_h_8es}} 
\caption{Cooling rate in $T$-$h$ plane, for (a) 2CS, (b) 2FS, (c) 4CS, (d) 4ES, (e) 5CS, and (f) 8ES clusters.} 
\label{fig:cool_rate}
\end{figure}

\begin{figure}[t!]   
\centering 
\subfigure{\includegraphics[scale=0.32,clip]{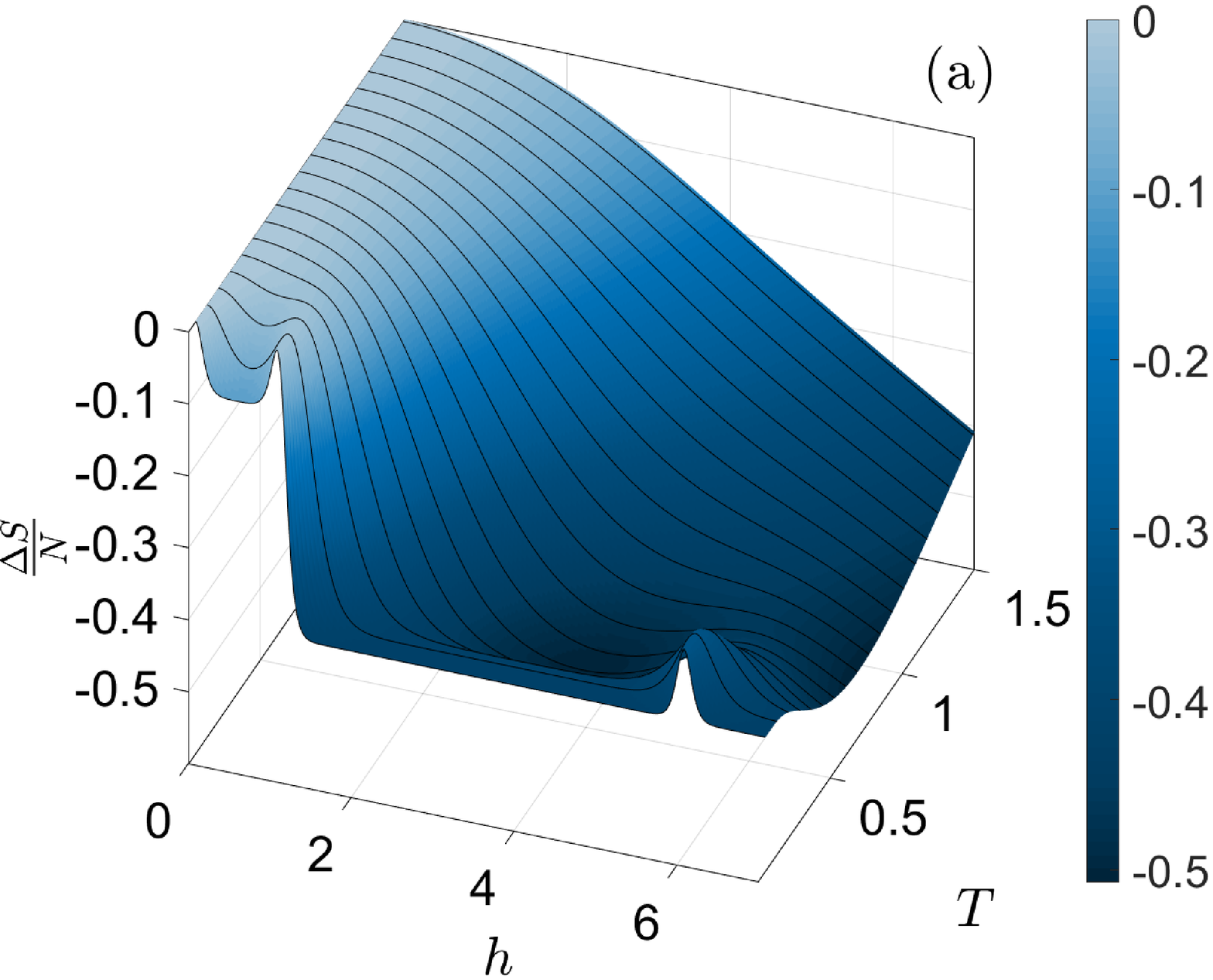}\label{fig:ent_side_2cs}} 
\subfigure{\includegraphics[scale=0.32,clip]{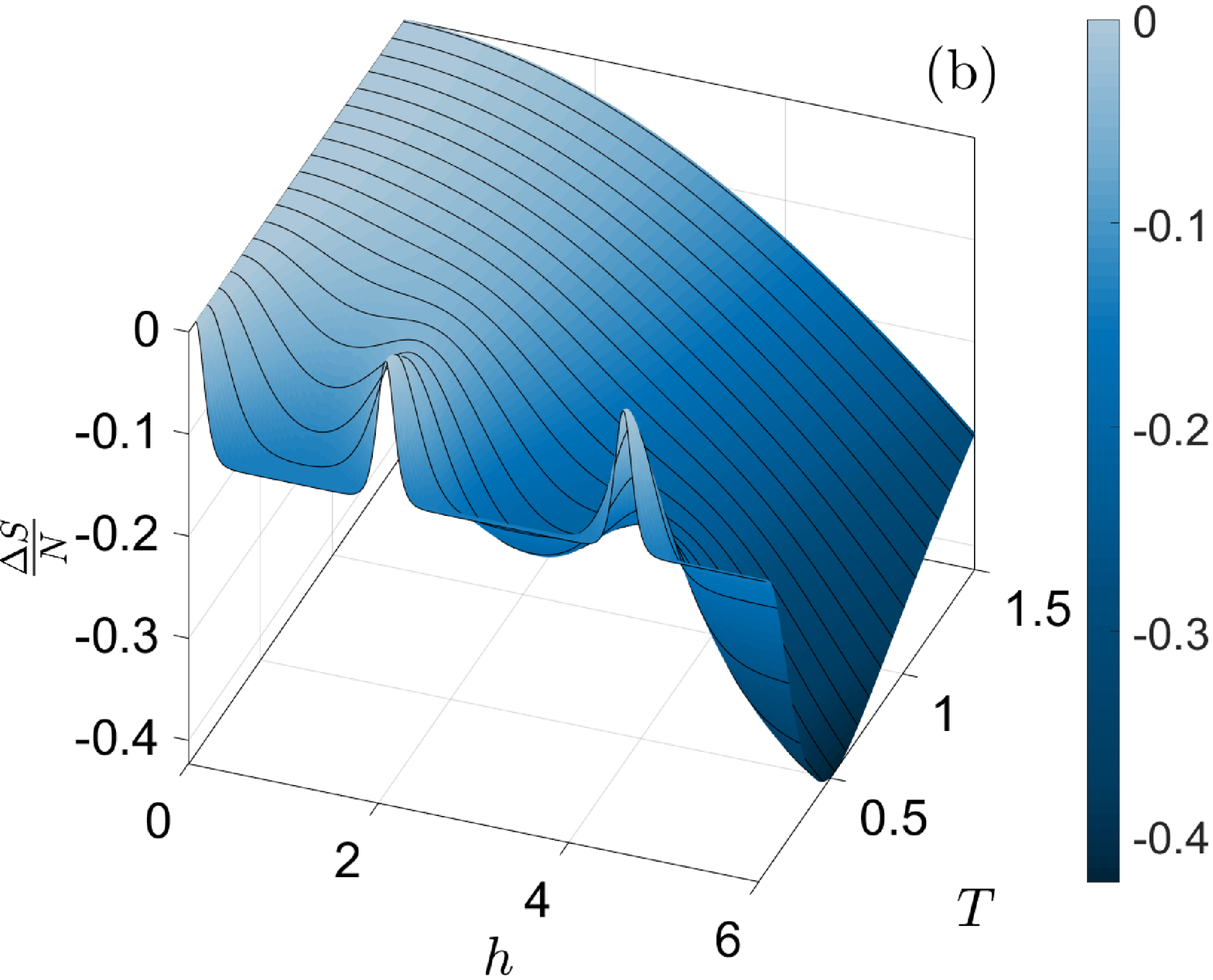}\label{fig:ent_side_2fs}}\\
\subfigure{\includegraphics[scale=0.32,clip]{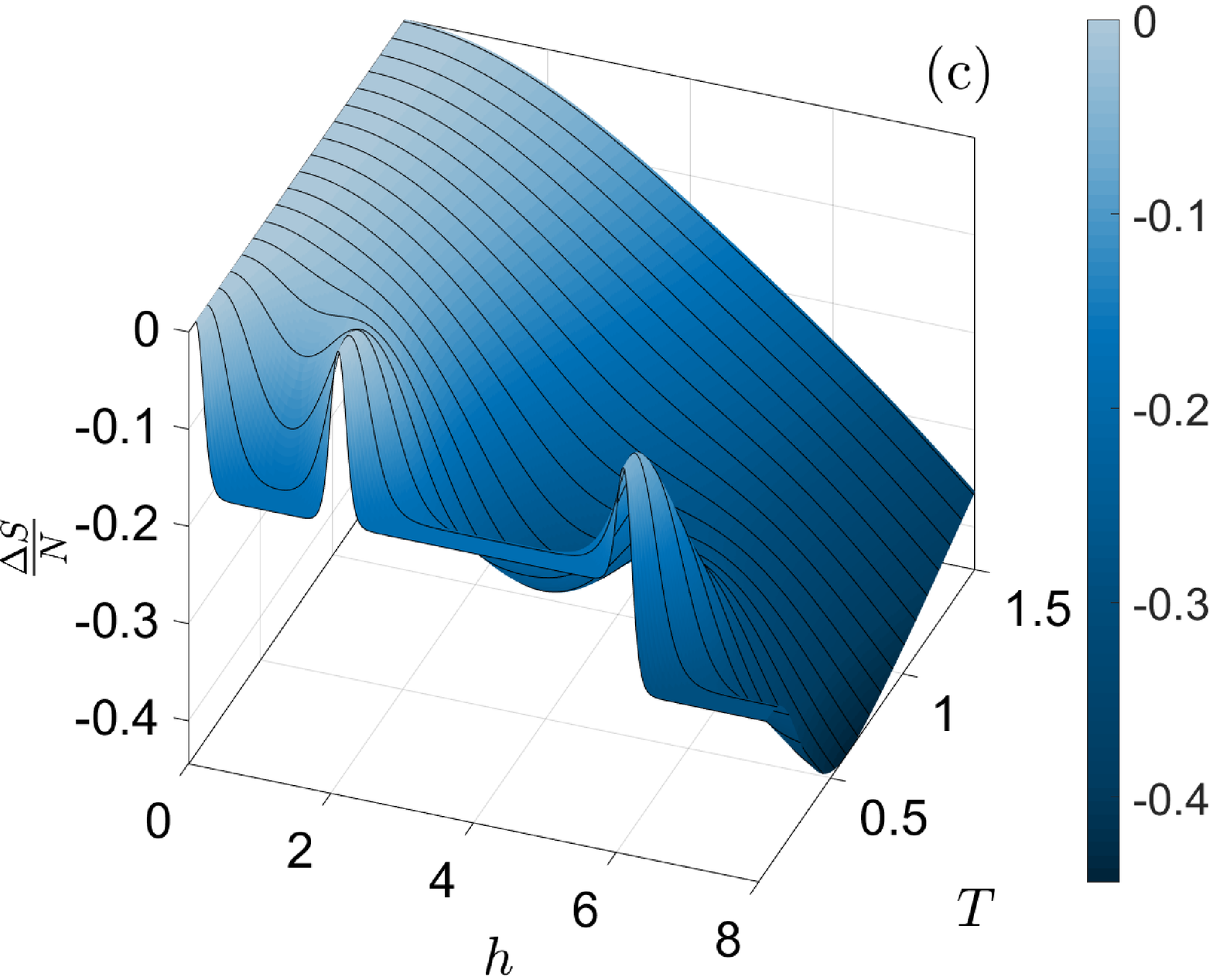}\label{fig:ent_side_4cs}}
\subfigure{\includegraphics[scale=0.32,clip]{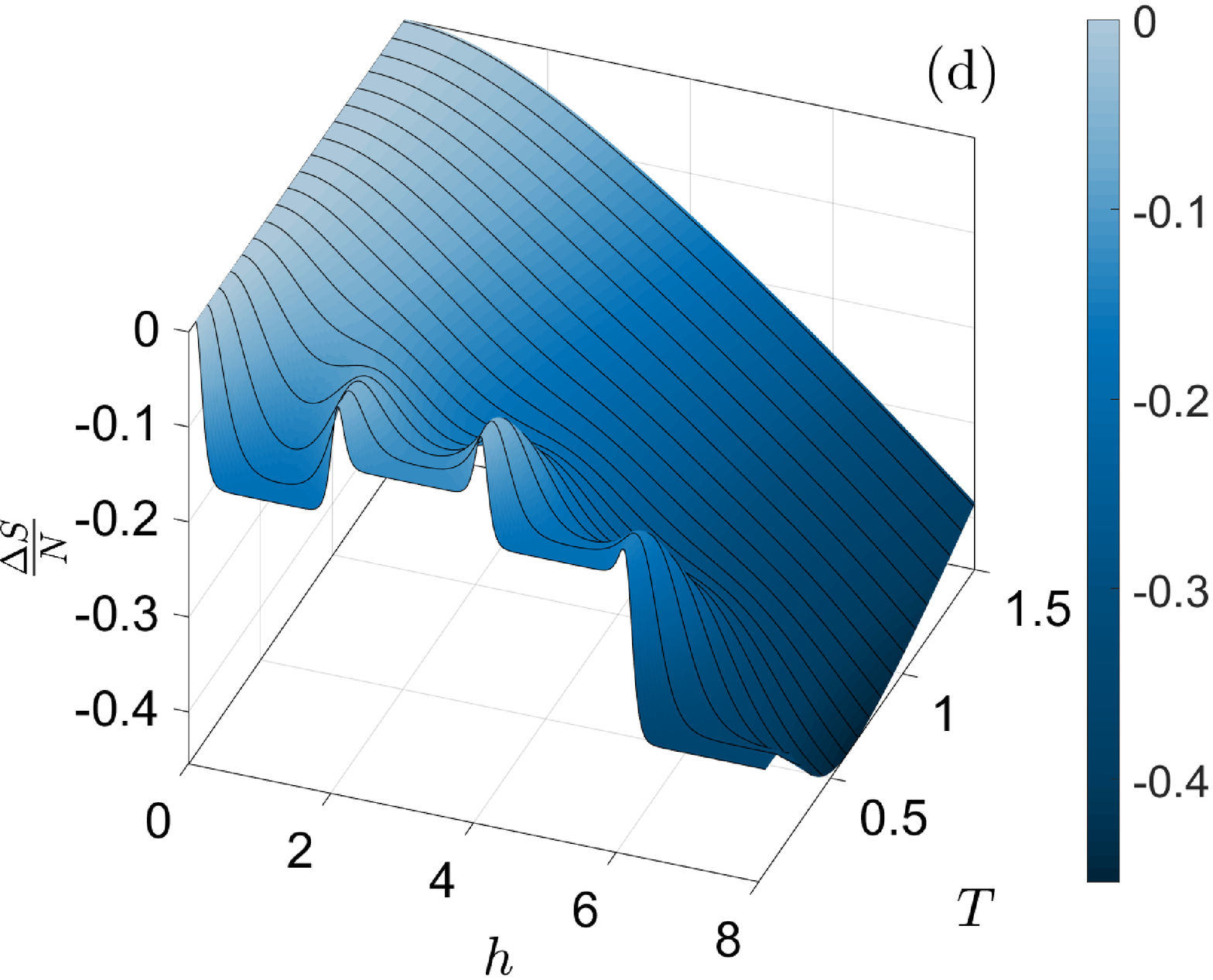}\label{fig:ent_side_4es}}\\
\subfigure{\includegraphics[scale=0.32,clip]{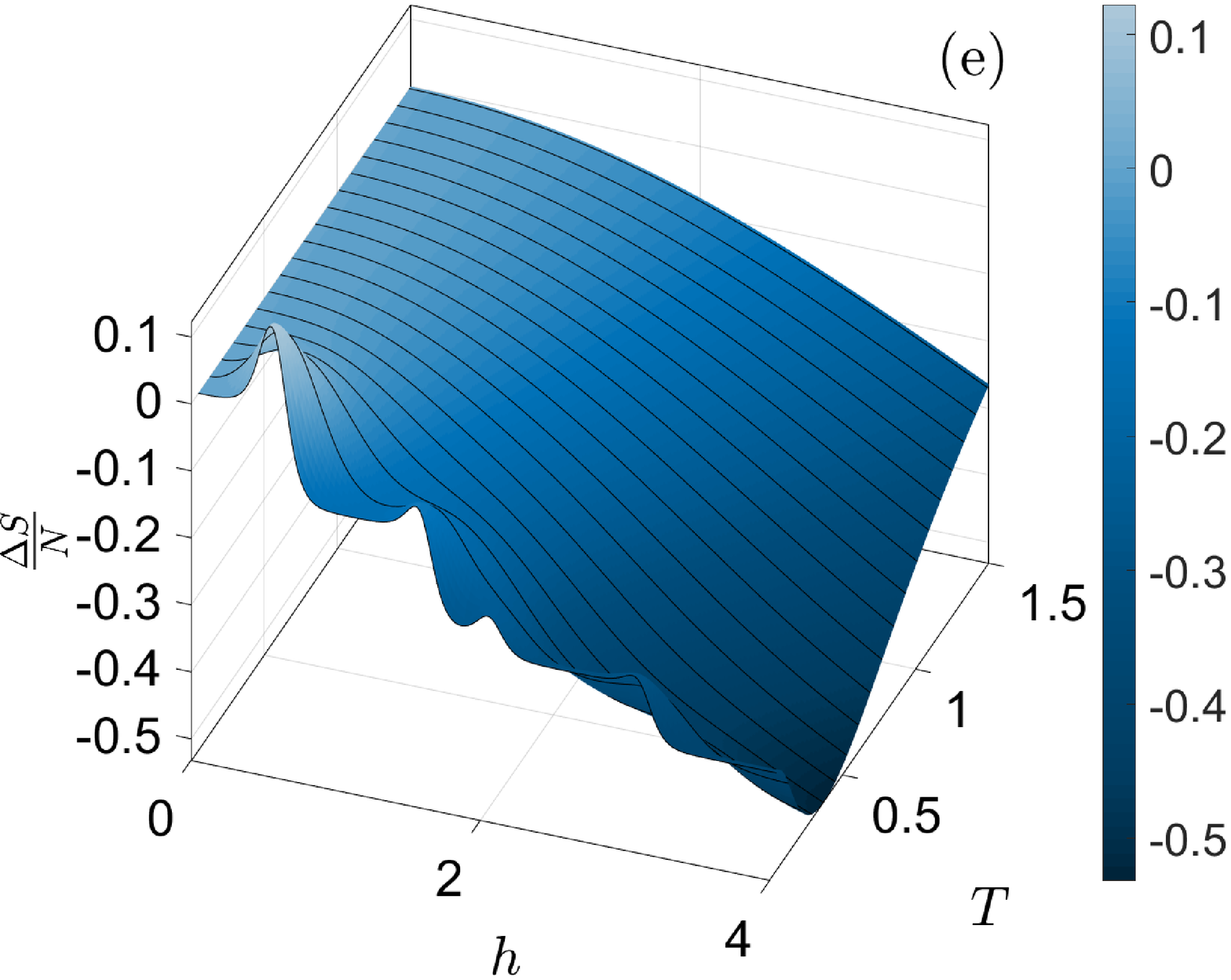}\label{fig:ent_side_5cs}}
\subfigure{\includegraphics[scale=0.32,clip]{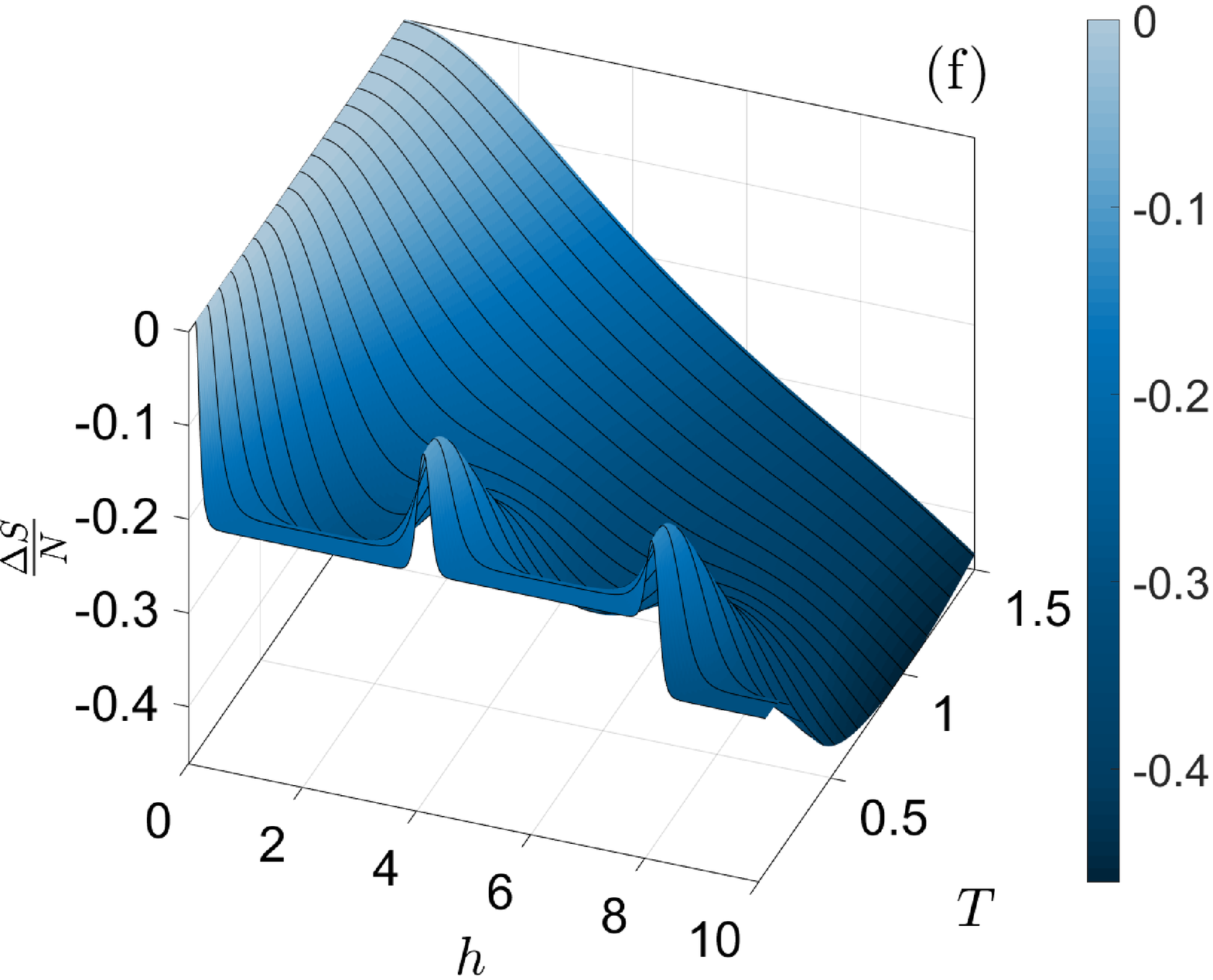}\label{fig:ent_side_8es}} 
\caption{Entropy density change in $T$-$h$ plane, for (a) 2CS, (b) 2FS, (c) 4CS, (d) 4ES, (e) 5CS, and (f) 8ES clusters.} 
\label{fig:ent_change_fin}
\end{figure}

As the plots of the magnetization in the $T$-$h$ parameter plane shown in Fig.~\ref{fig:mag_fin} demonstrate, the sharp step-wise variations with the increasing field, observed at zero temperature, get gradually smeared out by thermal fluctuations as the temperature increases. At the same time larger and larger fields are required to bring the respective systems to the saturated phase with all spins fully aligned to the field direction. \\
\hspace*{5mm} The dramatic low-temperature changes in both the magnetization and the entropy are interesting from the perspective of the magnetocaloric effect. In Fig.~\ref{fig:ent_fin} we present the top view of the latter in the $T$-$h$ plane for the respective clusters. The insentropes on the entropy density surface show temperature changes with a varying magnetic field under adiabatic conditions. One can see that the most prominent temperature responses occur in the vicinity of the critical fields at which the magnetization jumps between different plateaux. In particular the insentropes corresponding to the ground-state residual entropies at the respective critical fields display the largest gradients. For example, the 2CS cluster, the insentropes of which are presented in Fig.~\ref{fig:ent_top_2cs}, shows an enhanced magnetocaloric effect in the vicinity of the critical fields $h_{c1}=0$, $h_{c2}=1$, and $h_{c3}=6$ for the entropy density values close to $S_{GS}/N(h_{c1}=0)=(1/7)\ln(18) \approx 0.4129$, $S_{GS}/N(h_{c2}=1)=(1/7)\ln(16) \approx 0.3961$, and $S_{GS}/N(h_{c3}=6)=(1/7)\ln(2) \approx 0.0990$, respectively.\\
\hspace*{5mm} We note that, compared to the simple tetrahedron as well as the pyrochlore structure with zero magnetization persisting up to $h_{c1}=1$~\cite{stre15} and $h_{c1}=2$~\cite{jurc17}, respectively, most of the present tetrahedra-based finite clusters (all except 5CS) lack the zero-magnetization plateau. This feature allows for an enhanced magnetocaloric effect in the adiabatic demagnetization process at the entropy densities corresponding to the first critical field ($h_{c1}=0$) accompanied with a drastic (with infinite slope) decrease of temperature in a vanishing magnetic field. Similar magnetocaloric behavior has also been reported in some other geometrically frustrated Ising spin clusters, such as octaherdon and dodecahedron from the Platonic solids~\cite{stre15} or a nanocluster with a 'Star of David' topology~\cite{zuko18}.\\
\hspace*{5mm} The cooling rates in the respective clusters obtained in the entire $T$-$h$ parameter plane are presented in Fig.~\ref{fig:cool_rate}. In line with our expectations, based on the results presented above in Fig.~\ref{fig:ent_fin}, in absolute values the largest cooling rates correspond to the filed values in the vicinity of the critical fields in the respective clusters and low temperatures. In particular, large positive (negative) values are obtained just above (below) the critical fields, which correspond to the rapid cooling (heating) of the system. Except the 2FC cluster, in all the clusters in the limiting cases of $T \to 0$ and $h \to h_{ci}$ the cooling rates even diverge, which means the fastest possible cooling process. In the clusters 2CS, 4CS, 4ES and 8ES such a giant MCE can be achieved near zero magnetic field, which is of importance from technological point of view in effort to efficiently reach ultra-low temperatures by applying small magnetic fields.\\
\hspace*{5mm} Finally, in Fig.~\ref{fig:ent_change_fin} we present the entropy response to the change of the external field between different values. More specifically, the displayed quantities represent isothermal entropy density changes, as defined in Eq.~(\ref{entr_change}), with the initial field fixed to $h_i=0$. In all the structures, except the 5CS cluster, the entropy takes the maximum value in zero field and, therefore, for $h_i=0$ we observe a direct MCE. On the other hand, in the 5CS cluster an inverse MCE is possible for $h_f \approx 1$. The most rapid isothermal entropy density changes are observed at low temperatures and close to the critical fields. In particular, the largest entropy changes of about $\Delta S/N \approx -0.4$ can be achieved at relatively small fields of $h_f > 1$ in the 2CS cluster (see Fig.~\ref{fig:ent_side_2cs}).

\section{Conclusions}

We studied magnetic and magnetocaloric properties of various geometrically frustrated Ising antiferromagnetic nanoclusters assembled from tetrahedral units, as basic building blocks, which can share their vertices, edges or faces. It turned out that the topology of such clusters can significantly affect their properties, which can be very different from those of either a single tetrahedron~\cite{stre15} or the well studied infinite-lattice pyrochlore structure built of corner-sharing tetrahedra~\cite{jurc17,jurc14}. More specifically, the low-temperature magnetization process of the latter involves two plateaux at zero and one half of the saturation magnetization, separated by two critical fields. On the other hand, the present nanoclusters can feature from two plateaux with three critical fields up to four plateaux with four critical fields. \\
\hspace*{5mm} From the point of view of the magnetocaloric behavior an important difference lies in the fact that five out of the six studied clusters lack the zero-magnetization plateau, which is present in both the single tetrahedron and the pyrochlore lattice. The absence of the zero-magnetization plateau can facilitate a giant MCE in the form of a dramatic decrease in temperature in the vanishing magnetic field when the system is adiabatically demagnetized. Consequently, such systems might be suitable for technological application as efficient refrigerators to ultra-low temperatures. From this perspective, the nanocluster formed by two corner-sharing tetrahedra (2CS) appears to be the most promising. The lack of zero-magnetization plateau and a relatively short length of the first non-zero-magnetization plateau within $h_{c1}=0$ and $h_{c2}=1$ means that the area of an increased cooling rate observed in the vicinity of the critical fields is concentrated at relatively small field intensities. Furthermore, beyond $h_{c2}=1$ the ground-state degeneracy is completely lifted, which in combination with a relatively large zero-field GS degeneracy leads to fairly large isothermal entropy density changes upon applying relatively small fields. \\
\hspace*{5mm} In the future, we believe that it could be interesting to study effects of certain parameters, such as exchange and single-ion anisotropies in selected nanoclusters with larger spin values. We expect, that besides the nanocluster topology also these parameters can be used to further tune magnetocaloric properties with the goal to look for the best candidates to be used in practice as efficient magnetic coolers.

\section*{Acknowledgments}
This work was supported by the grant of the Slovak Research and Development Agency under the contract No. APVV-16-0186 and the Scientific Grant Agency of Ministry of Education of Slovak Republic (Grant No. 1/0531/19).

\end{document}